\documentclass[letter]{aa} 
\usepackage{natbib,twoopt}
\usepackage{titlesec} \usepackage{amsmath, amssymb}
\usepackage{txfonts}
\usepackage[normalem]{ulem}
\usepackage{graphicx}
\usepackage{comment}

\usepackage{hyperref}
\newcommand{\Teff}{${T}_{\rm eff}$}
\newcommand{\logg}{$\log g$}

\begin{document}

   \title{Detectability limits of Ce abundances in Gaia RVS}


   \author{S. Vitali
          \inst{1,2}
          \and
          A. Escorza \inst{1,2}
          \and
          T. Signor
          \inst{3}
          \and
          P. A. Palicio\inst{1,2,4}
          \and
         M. Abdul-Masih\inst{2,1}
        \and
          C. Allende Prieto\inst{1,2} 
          \and
          D. Godoy-Rivera\inst{1,2}
        \and
         J. I. González Hernández\inst{1,2}
          \and
          M. Schultheis\inst{4}
          \and
    S. Shetye\inst{5,6} 
    \and
     S. Van Eck\inst{5,7} 
          }
   \institute{Instituto de Astrofísica de Canarias, C. Vía Láctea, s/n, 38205 La Laguna, Santa Cruz de Tenerife, Spain.\\
    \email{svitali@iac.es}
      \and
    Universidad de La Laguna, Dpto. Astrofísica, Av. Astrofísico Francisco Sánchez, 38206 La Laguna, Santa Cruz de Tenerife, Spain.\
    \and
    Instituto de Estudios Astrofísicos, Facultad de Ingeniería y Ciencias, Universidad Diego Portales, Av. Ejercito 441, Santiago, Chile.\
    \and
    Universit\'e C\^ote d'Azur, Observatoire de la C\^ote d'Azur, Laboratoire Lagrange, CNRS, Blvd de l’Observatoire, 06304 Nice, France.
    \and
    Institute of Astronomy and Astrophysics, Université libre de Bruxelles, CP 226, Boulevard du Triomphe, 1050 Bruxelles, Belgium.
    \and
    Institute of Astronomy, KU Leuven, Celestijnenlaan 200D, 3001 Leuven, Belgium.
    \and
      BLU-ULB, Brussels Laboratory of the Universe, blu.ulb.be, Brussels, Belgium.
    }
 \date{Received \today}

\abstract
  {\textit{Gaia} DR3 provides medium-resolution Radial Velocity Spectrograph (RVS) spectra for millions of stars, but the extent to which they contain recoverable cerium (Ce) information across stellar parameter space remains unclear.}
  {We assess Ce detectability in \textit{Gaia} RVS spectra and the reliability of DR3 Ce abundances ahead of the larger DR4 catalogue.}
  {We compute synthetic spectra across a stellar-parameter grid representative of the \textit{Gaia} RVS sample. We combine the Cayrel equivalent-width criterion and the Cramér-Rao bound to map Ce detectability as a function of stellar parameters and signal-to-noise (SNR). We also quantify continuum-placement errors and compare the predicted detectability boundary with the \textit{Gaia} DR3 GSP-Spec catalogue.}
  {Ce is reliably detectable primarily in cool giants at [Fe/H]$\gtrsim-0.5$~dex and with at least [Ce/Fe] $\gtrsim0$ at an average SNR of 50, and the detectability region expands toward higher gravities at increasing SNR. Continuum-placement errors of $\gtrsim1\%$ become a significant source of bias, and at $\gtrsim2\%$ they completely prevent Ce detection.}
  {The criterion we present identifies where RVS Ce abundances, including those from DR4, can be robustly exploited. Outside our detectability region, these abundances should be interpreted cautiously, whether derived by GSP-Spec or by supervised methods. Currently, depending on the assumed continuum-placement uncertainty, between $\sim$40\% and $\sim$70\% of the published DR3 Ce measurements fall within this region; however, the rest are not reliable. Our findings are particularly important for metal-poor stars, for which the Ce lines in the RVS range are simply not detectable.}


   \maketitle
%

\section{Introduction}

Our understanding of the Milky Way and its stellar populations has advanced dramatically in recent years, driven by the unprecedented volume of observational data now available. The ESA \textit{Gaia} mission \citep{2016Gaia} has delivered astrometric and photometric information for $\sim$1~billion stars \citep[\textit{Gaia}~DR3,][]{2023Gaiacoll} including stellar parameters and chemical abundances for about 5.6~million stars through the General Stellar Parametrizer from Spectroscopy (GSP-Spec) module \citep[hereafter RB23]{2023Recio}, based on medium-resolution spectra (\textit{R}$\sim$11,500) obtained with the \textit{Gaia} Radial Velocity Spectrometer \citep[RVS;][]{2018Cropper,2023Katz}. Among the abundances provided by GSP-Spec in \textit{Gaia} DR3, cerium (Ce) is one of the neutron(n)-capture elements available. This group of elements is particularly valuable as a tracer of distinct enrichment histories \citep{2003Sneden} and, in chemically peculiar stars, of past binary mass-transfer events \citep{1980McClure}. Specifically, Ce, an s-process element of the second peak, has been used to map n-capture abundance patterns across the Galactic disc and halo \citep{2023Contursi}, probe AGB nucleosynthesis sites \citep{2024Contursi}, and identify candidate barium-star binaries \citep{2025Navarrete}. However, weak and often blended spectral features make n-capture elements particularly challenging to recover from medium-resolution data, raising questions about the reliability of the measurements. In \textit{Gaia} RVS  spectra, Ce abundances are derived from a \ion{Ce}{II} triplet around 851.1\,nm, which is unresolved at the RVS resolving power. Moreover, for certain combinations of atmospheric parameters, this triplet can be blended with a weak CN feature near 851.25~nm \citep{2023Recio,2023Contursi}. 

These limitations will become increasingly relevant with the arrival of \textit{Gaia}~DR4, which is expected to deliver GSP-Spec chemical abundances for over 36~million sources, expanding the Ce catalogue and exacerbating the reliance on automated selection and validation methods. In this context, it is essential and timely to investigate whether the Ce feature at RVS resolution carries enough spectral information to serve as a reliable diagnostic, particularly since these measurements are already being used as input to automated, machine learning (ML)-based selection methods\citep{2026Das}. 

In this Letter, we investigate this by applying a detectability criterion on synthetic spectra (Sect.~\ref{sect:ce_sensitivity}), performing a comparison with the \textit{Gaia}~DR3 Ce catalogue (Sect.~\ref{sect:comparison}), and discussing the implications of our findings for population studies (Sect.~\ref{sect:studies}). 

\section{Spectral Sensitivity Analysis}\label{sect:ce_sensitivity}

To assess the spectral information on Ce in \textit{Gaia} RVS  spectra, we first define our Ce detectability criterion. Then, we take into account observational effects by examining its sensitivity to SNR and continuum normalization errors. We begin by synthesizing a grid of 1\,588\,600 spectra over the 851-851.6 nm range, covering the following parameter space: \Teff = 3700 to 6025 K (step 25 K), $\log g = $ 0.0 to 4.8 dex (step 0.2 dex), [Fe/H] = -2.0 to +0.5 dex (step 0.1 dex), and [Ce/Fe] = -0.5 to +2.0 dex (step 0.1 dex). A resolving power of $R$=11,500 was adopted to match the \textit{Gaia} RVS. The synthesis setup is described in Appendix~\ref{append:synthesis}.

\subsection{Ce detectability}\label{sect:detectability}
We assess the spectral information available for Ce measurements using the Cramér-Rao bound \citep[CRB,][]{Cramer1945,Rao1945} computed from the spectral gradient with respect to [Ce/Fe]. This allows us to evaluate the sensitivity of a spectrum to variations in a specific stellar label \citep{2017Ting}. We define Ce as reliably detectable when the achievable precision satisfies $\sigma_{\rm{[Ce/Fe]}}<0.1$ dex, a threshold comparable to typical uncertainties reported in high-resolution abundance studies of barium stars \citep[e.g.,][]{2026Escorza}. This theoretical lower bound assumes a locally linear response of the spectrum to [Ce/Fe] variations and Gaussian, uncorrelated pixel noise, and represents the best precision achievable when estimating [Ce/Fe] from the data. Its detailed derivation and justification are given in Appendix~\ref{appendix:gradient}. 
\begin{figure}[!t]
    \centering   \includegraphics[width=\columnwidth]{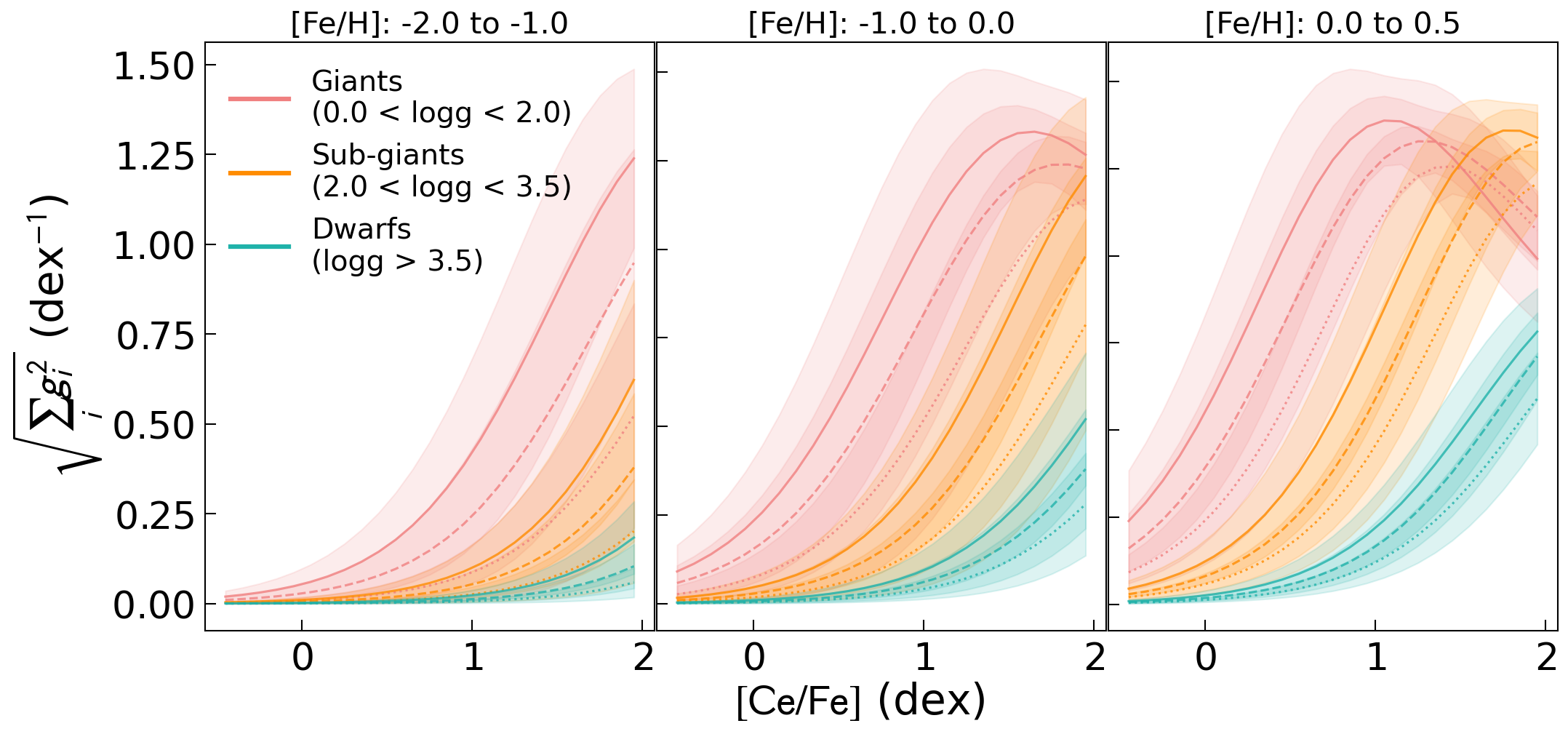}
    \caption{Root-sum-square of the per-pixel spectral gradient across the Ce window vs. [Ce/Fe] for three metallicity bins, spanning the full range of atmospheric parameters. The three lines represent various temperature classes: namely, 3700–4500 K (solid line), 4500–5300 K (dashed line), and 5300–6025 K (dotted line). Shaded regions illustrate the $\pm1\sigma$ dispersion around the mean gradient.}  \label{fig:gradient_bin}
\end{figure}

The spectral response to Ce variation is illustrated in Fig.~\ref{fig:gradient_bin} as $\sqrt{\sum_i g_i^2}$ vs [Ce/Fe], where $g_i = \partial f_i/\partial [\mathrm{Ce/Fe}]$ (see Eq.~\ref{eq:cramer}), across the full grid, computed within the abundance window indicated by RB23 (see their Table B.1). The spectral response to Ce increases monotonically with [Ce/Fe] in giants, approximately linearly up to $\sim$1.0 dex, before saturating. This saturation is strongest in the metal-rich bin and absent in the metal-poor bin, where the gradient is still rising at [Ce/Fe]$\sim$2~dex. For sub-giants and dwarfs, the gradient is much weaker but remains approximately linear across the plotted range, since the line is too shallow to saturate. Only at higher abundances ([Ce/Fe] $\gtrsim 1.0$ dex) does the gradient begin to saturate, particularly in metal-rich bins.  
These trends validate the locally linear assumption embedded in the CRB formula (Eq.~\ref{eq:cramer}), over the regime relevant to our detectability analysis ([Ce/Fe] from 0 to +0.5~dex).

While the CRB incorporates photon noise through the assumed SNR, it combines the signal in quadrature over all pixels within the Ce line window. As a result, a line may be statistically detectable even when its core depth is only a small fraction of the per-pixel noise, a regime that is highly sensitive to systematics such as continuum placement and blending. This bound therefore provides an optimistic detectability threshold for the true detectability. To ensure that the predicted sensitivity corresponds to a genuinely measurable spectral feature, we complement this bound with a second requirement that the Ce line core be individually visible above the noise, following the standard detection criterion of \cite{1988Cayrel}. 
We compare the measured equivalent width of the line, integrated directly against a local linear continuum defined from the edges of the Ce window, to its Cayrel uncertainty, $\sigma_{W} = \frac{1.5}{\mathrm{SNR}}\sqrt{\mathrm{FWHM}
\cdot d_{x}}$, and require a $3\sigma$ detection for the equivalent-width (EW), $\mathrm{EW}\geq3\sigma_{W}$. The
synthetic Ce line width varies little across dwarfs, giants, and metallicity
in our grid ($0.086\lesssim\mathrm{FWHM}\lesssim0.10$~nm), so we adopt the intermediate value, $\rm{FWHM}\approx0.094$~nm, while $d_{x}=0.03$~nm is the pixel size (RB23).

\begin{figure}[t]
    \centering   \includegraphics[width=\columnwidth]{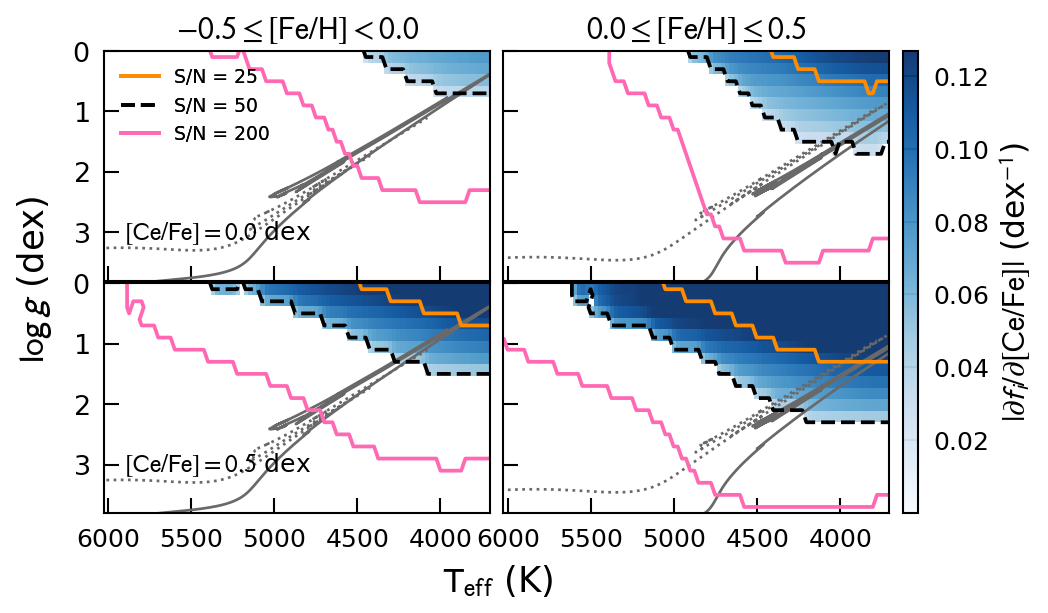}
    \caption{Spectral sensitivity to Ce variations evaluated at [Ce/Fe]=0.0; 0.5~dex in the Kiel diagram for two metallicity bins. Contours mark the detectability threshold ($\sigma_{\rm[Ce/Fe]}<0.1$) at different SNR. Solid and dotted curves show the 1;$2\,\mathrm{M_\odot}$ MIST evolutionary tracks \citep{2016Choi,2016Dotter}, computed for [Fe/H] = $-0.5;+0.5$ (1 $\mathrm{M_\odot}$) and [Fe/H] = $-0.25;+0.25$ (2 $\mathrm{M_\odot}$) in the left and right panels, respectively.}  \label{fig:kiel_sensitivity}
\end{figure}

\subsubsection{Sensitivity to signal-to-noise ratio}\label{sect:snr}

Fig.~\ref{fig:kiel_sensitivity} shows the Ce gradient across the Kiel diagram for the metallicity and [Ce/Fe] bins where the line satisfies both the CRB precision and the EW (Cayrel) detection criteria (Sect.~\ref{sect:detectability}).

To assess the impact of SNR on Ce detectability, we adopted three representative values. $\rm{SNR}=25$, lies just above the $\rm{SNR}>20$ threshold used by GSP-Spec for RVS parametrizationin RB23 and is comparable to the regime targeted by supervised ML RVS methods such as in \cite{2026Das}. $\rm{SNR}=50$, corresponds to the more stringent \textit{Gaia} RVS  threshold adopted by \cite{2026Das} for reliable stellar parameters and chemical abundances. $\rm{SNR}=200$, matches the quality threshold adopted by \cite{2023Creevey} for reliable astrophysical parameters. Since both the CRB precision and the EW detection threshold scale as $1/\mathrm{SNR}$, we rescale them directly from the intermediate value,
$\rm{SNR}=50$ (black dashed line in Fig.~\ref{fig:kiel_sensitivity}), to $\rm{SNR}=25$ and $200$ (orange and pink contours,
respectively). 

At sub-solar metallicity (left panels) and [Ce/Fe]~$=0.0$, $\rm{SNR}=25$ fails to yield a detectable region, while at [Ce/Fe]~$=+0.5$ all three SNR values are sufficient, and the detectable region extends 
almost to \logg~$\simeq3$ for the highest SNR. This effect is even more pronounced at
near- and super-solar metallicity (right panel), where the detectable region
reaches \logg~$\sim3~$--$~4$, i.e. into the subgiant and dwarf regime, at $\rm{SNR}=200$. However, even for [Ce/Fe]~$=+0.5$, the detectable area contracts sharply at $\rm{SNR}=25$, indicating that a non-negligible fraction of RVS spectra fall below the SNR required to obtain reliable Ce measurements, even in a favourable metallicity regime. 

Extending to lower metallicities (Appendix~\ref{appendix:spectral},
Fig.~\ref{fig:kiel_sensitivity_all}), the Ce sensitivity at [Fe/H]~$<-0.5$ is
strongly [Ce/Fe]-dependent. At solar or sub-solar [Ce/Fe], the line remains essentially undetectable except in the coolest giants even at SNR$=200$. For [Ce/Fe]~$\gtrsim0.5$~dex, the detectable region grows but remains confined to cool giants. At $\rm{SNR}=50$ it reaches \logg~$\sim1~$--$~2$ (depending on enhancement), extending only to \logg~$\sim2.5~$--$~3$ at $\rm{SNR}=200$. The Ce abundance diagnostic is therefore useful in this metallicity regime only for sufficiently enhanced, evolved, high-SNR targets and is not applicable to dwarfs.

\subsubsection{Sensitivity to normalization}\label{sect:continuum}

Our analysis assumes a perfectly known continuum, but continuum-placement errors are an additional source of uncertainty for weak features such as Ce \citep{2010Pancino,2017Jofre}. Given the narrow extent of our gradient window ($\approx0.2$ nm),  we approximated the normalization uncertainty as a constant offset $\epsilon$, consistent with the GSP-Spec normalization strategy \citep[for more details, see][]{2020Santos,2023Recio}. Under the local linearity assumed by the CRB, this offset translates into a spurious abundance shift $\Delta\mathrm{[Ce/Fe]} \approx \epsilon / (\partial f/\partial \mathrm{[Ce/Fe]})$. In Appendix~\ref{append:norm}, a detailed derivation is given together with the bias induced across the Kiel diagram (see Fig.~\ref{fig:continuum}).

We found that, for $-0.5<\mathrm{[Fe/H]}<+0.5$ dex and [Ce/Fe]~$=+0.5$ dex, the induced bias remains below 0.1~dex for $\epsilon=0.2\%$, exceeds this threshold over $\sim$20\% of the detectable region for $\epsilon=1\%$, and dominates the uncertainty throughout the detectable region for $\epsilon\gtrsim2\%$. Continuum-placement errors should therefore be considered alongside the CRB when assessing where Ce can be reliably recovered from RVS spectra.

\section{Comparison with GSP-Spec measurements}\label{sect:comparison}
\begin{figure}[t]
    \centering   \includegraphics[width=0.5\textwidth]{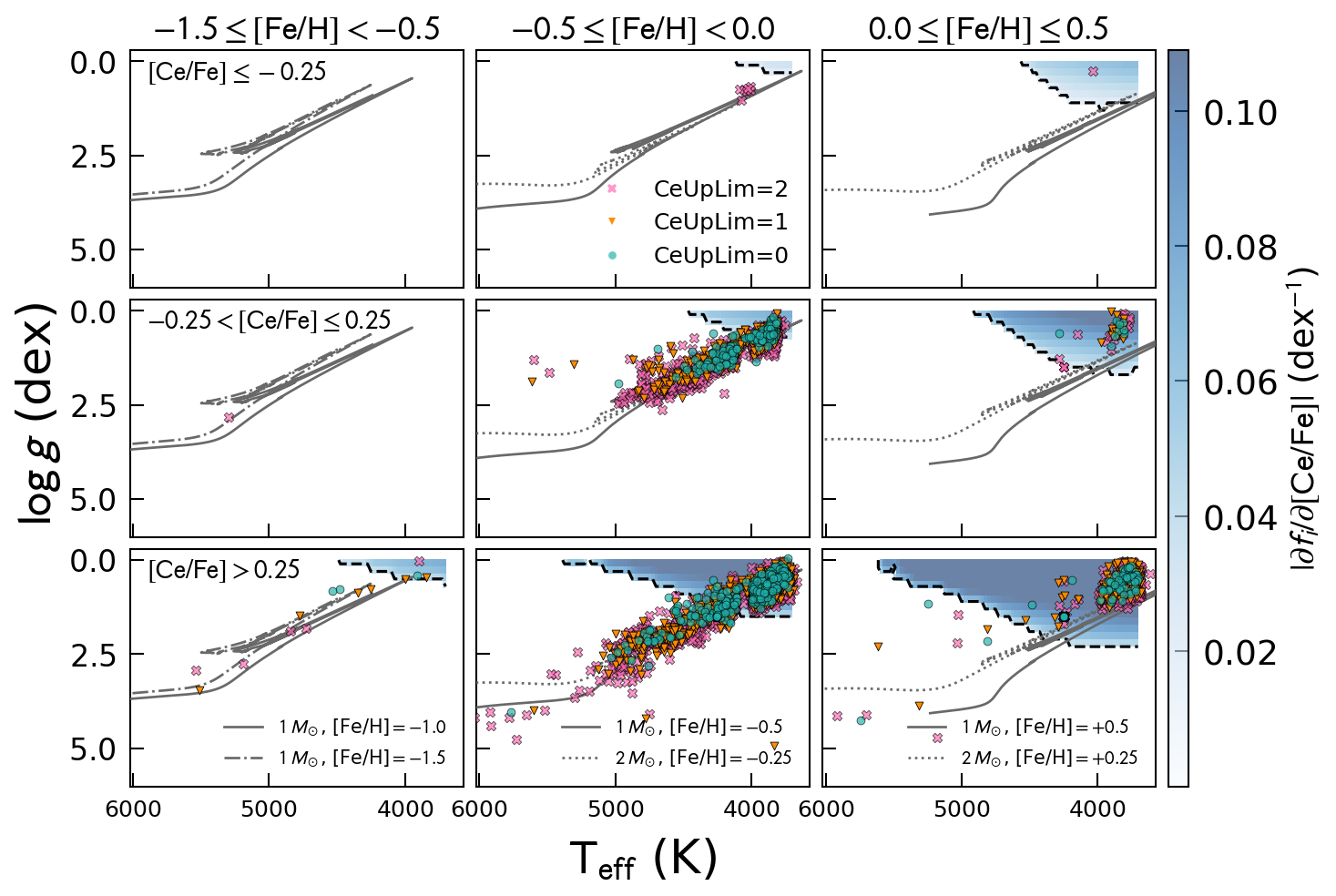}
    \caption{Ce detectability boundary (black dashed line, CRB+EW criterion at SNR$=50$) compared with Gaia DR3 stars, binned by [Fe/H] and by each star's own [Ce/Fe]. Points are coloured and
marker-coded by GSP-Spec quality flag. Same evolutionary tracks as in Fig.~\ref{fig:kiel_sensitivity}, with the addition of the $1\,\rm{M_{\odot}}$ tracks at [Fe/H]$=-1.5$ and $-1.0$ (first column).}  \label{fig:gaia_sample}
\end{figure}

To place our theoretical detectability boundary in the context of the actual Ce measurements delivered by GSP-Spec, we examine how the current Gaia DR3 Ce catalogue is distributed relative to it. Restricting the sample to stars with $\rm{SNR}\geq50$, non-null \Teff\ and \logg, a broadening parameter $v_{\rm broad} \leq15,\rm{km},\rm{s}^{-1}$, and a measured Ce abundance with  \texttt{CeUpLim}~$\leq2$\footnote{\texttt{CeUpLim} indicates the Ce line depth relative to the noise level and flags secure (0), moderately confident (1), or uncertain (2) measurements, following RB23 (see their Sect.~8.7 and Tables~C.9 and C.11)}, we retain $9930$ stars. We bin them by their measured [Ce/Fe], assigning each to the closest of our three reference values ($-0.5$, $0.0$, $+0.5$~dex) i.e. splitting at the midpoints $[\mathrm{Ce/Fe}]=-0.25$ and $+0.25$. 

Fig.~\ref{fig:gaia_sample} shows these stars overlaid on our theoretical boundary (CRB+EW, computed at each row's reference [Ce/Fe]) and coloured by the GSP-Spec quality flag from RB23. The resulting distribution is strongly skewed toward enhanced abundances: only $8$ stars ($0.1\%$) fall in the [Ce/Fe]~$\leq-0.25$ bin, $18.0\%$ in the range $-0.25<\mathrm{[Ce/Fe]}\leq0.25$, and the majority, $81.9\%$,
in [Ce/Fe]~$>0.25$, itself consistent with our prediction that Ce is essentially undetectable at low or negative [Ce/Fe] value except in the coolest
giants (Sect.~\ref{sect:snr}).

The best-quality
measurements (\texttt{CeUpLim}~$=0$) concentrate densely within the predicted detectability region, clustering along its cool, low-\logg\ edge,
particularly in the metal-rich, enhanced-[Ce/Fe] panels where our
criterion predicts the most favourable detectability.
In the intermediate-metallicity bin ($-0.5<\mathrm{[Fe/H]}<0$), however, $\gtrsim$30\% of \texttt{CeUpLim}~$=0$ stars (and a larger fraction of \texttt{CeUpLim}~$=1-2$) fall outside our boundary, consistent with these being GSP-Spec's least confident tiers, where empirical line-depth-to-noise assessments and our synthetic-grid criterion are expected to agree least well. This scatter is unsurprising given our conservative comparison: our boundary assumes a perfectly known continuum and fixed atmospheric parameters, neglects continuum bias
(Sect.~\ref{sect:continuum}) and uncertainties in \Teff\ and \logg, and it is derived from synthetic spectra rather than a direct assessment of the GSP-Spec pipeline output. Even so, the close correspondence between our boundary and the location of the highest-quality measurements supports its use as a practical guide to decide where Ce abundances can be reliably recovered from RVS spectra.

\section{Implications for stellar population studies}\label{sect:studies}

Our synthetic-spectra analysis (Sect.~\ref{sect:ce_sensitivity}) and comparison with the available \textit{Gaia} DR3 measurements (Sect.~\ref{sect:comparison}) indicate that at typical survey SNRs, Ce abundances from the \textit{Gaia} RVS can be reliably recovered only for cool giants ($\log g\lesssim2$) with $\mathrm{[Fe/H]}\gtrsim-0.5$~dex and at least solar-level Ce enhancement ($\mathrm{[Ce/Fe]}\gtrsim0.0$-$0.5$~dex, depending on metallicity). 

Combining our detectability criterion with the \texttt{CeUpLim} flag, a conservative sample for population studies requires $\texttt{CeUpLim}\leq2$, $\mathrm{[Fe/H]}\gtrsim-0.5$ dex, and $\mathrm{SNR}\gtrsim50$. Of the $\sim$9900 DR3 stars satisfying these criteria, $\sim$71\% ($\sim$6900) fall within our EW+CRB detectability region, decreasing to $\sim$40\% ($\sim$3800) if we consider a 1\% continuum-placement error (see Table~\ref{tab:ceuplim_stats} and Appendix~\ref{append:E} for a detailed breakdown by metallicity bin). 
With DR4 expected to parameterise around 36 million stars with RVS, and a gain factor in SNR of $\sim1.4$ due to the extra 32 months of observations compared to DR3, the fraction of detectable Ce measurements is likely to increase.

This is qualitatively consistent with \cite{2023Contursi}, who also found the Ce triplet easier to detect in cool giants than in dwarfs,  although with less restrictive limits  (\Teff$\sim5400$ K, \logg$\gtrsim3.5$) using a noise-independent criterion based on a 0.5\% line-core flux depression relative to a synthetic spectrum with [Ce/Fe]~$=-2.0$ dex. By neglecting noise, their criterion predicts a larger detectable region and is not directly comparable to ours.
This metallicity floor limits the use of \textit{Gaia} RVS Ce abundances as n-capture tracers. Although low-resolution \textit{Gaia} XP spectra can identify carbon-enhanced metal-poor (CEMP) star candidates from CH \citep[e.g.][]{2023Lucey,2025Arentsen}, RVS  spectra cannot provide reliable Ce abundances in the metal-poor regime where these stars are particularly important. Consequently, Ce cannot be used to infer complementary s-process information for the chemical characterization of these CEMP populations. 

This limitation also affects chemically peculiar stars such as barium stars, predominantly identified at $\mathrm{[Fe/H]}\gtrsim-0.5$~dex \citep{2016decastro,2026Escorza}, despite being, on average, more metal-poor than the field population \citep{2026Levine}.  Ce detectability may thus bias searches for Ce-enhanced stars towards higher metallicities, particularly for CEMP-s stars, whose s-process signatures are most relevant in the metal-poor regime where Ce is effectively undetectable in RVS spectra.

Supervised ML methods have also been used to infer n-capture abundances, including Ce, from RVS spectra by transferring labels from higher-resolution surveys \citep{2026Das}, potentially recovering information beyond a single feature criterion like ours, by exploiting correlations across the full spectrum. However, \cite{2026Das} find that Ce shows the largest scatter and the strongest degradation toward low SNR among their labels, consistent with a weak, marginal per-line signal.  This raises the question of whether predicted Ce abundances reflect genuine spectral information or learned correlations with other stellar parameters, in which case they may trace the star's position in the correlated parameter space rather than its true Ce content. Where our criterion predicts little detectable Ce signal, these abundances should not be interpreted as direct measurements without independent validation.

Validations by \cite{2023Recio} and \cite{2023Contursi} find good agreement with high-resolution measurements \citep{2019Forsberg} within, or close to, our favourable region, though \cite{2023Contursi} extend their measurements to lower metallicity ([Fe/H]$\gtrsim-1.0$) at much stricter SNR ($\geq300$) to reach the precision needed for
population-level Galactic trends. Agreement degrades specifically at low [Ce/Fe], where literature comparisons reveal systematic discrepancies. 

\section{Conclusions}
We assessed the detectability of Ce in \textit{Gaia} RVS  spectra using synthetic spectra, the CRB and equivalent-width criteria, and a comparison with the \textit{Gaia}~DR3 Ce catalogue. The code developed for this work is available on \url{https://github.com/Sara-Vitali/ce-rvs-detectability.git}. Our main results are as follows:
   \begin{enumerate}
      \item Ce is reliably detectable only in cool giants ($\log g\lesssim2$), at $\mathrm{[Fe/H]}\gtrsim-0.5$~dex, and with at least solar-level [Ce/Fe] enhancement, and it is essentially undetectable in dwarfs and metal-poor stars at the typical SNR in DR3.
     \item The detectable region expands toward warmer \Teff\ and higher gravities with increasing SNR, but continuum-placement errors above $\sim1\%$ become the dominant error source in the
    intermediate-metallicity regime.
    \item The majority of \textit{Gaia}~DR3 stars ($\sim 40\%-71\%$) with a high-quality Ce measurement fall within our predicted detectable region,  validating our criterion against the published DR3 quality flags.
   \end{enumerate}
These limits preclude studies of [Ce/Fe] in metal-poor stars (e.g., CEMP classification) or searches for n-capture/Ce-enhanced stars at $\mathrm{[Fe/H]}\lesssim-0.5$~dex. While we fully acknowledge the transformative potential of \textit{Gaia} data, with this Letter we aim to encourage careful use of Ce abundances from \textit{Gaia} RVS  spectra, particularly from ML-based methods, whose need for large training samples may push them toward low-SNR stars where the Ce triplet carries little or no detectable signal, and a predicted abundance may instead reflect correlations with other stellar parameters.
In this scenario, our detectability map also provides concrete guidance on where Ce measurements from the upcoming \textit{Gaia}-DR4 can be exploited, for instance to search for s-process-enhanced stars or to place Ce upper limits constraining n-capture enrichment in the parameter range where these features are genuinely informative. Finally, although we focus on Ce, the proposed methodology can be readily extended to other abundances derived by GSP-Spec, which we plan to explore in future work.

\begin{acknowledgements}
This research used the Astrophysics Data System (NASA Cooperative Agreement 80NSSC21M0056) and data from the ESA \textit{Gaia} mission (\url{https://www.cosmos.esa.int/gaia}), processed by the DPAC (\url{https://www.cosmos.esa.int/web/gaia/dpac/consortium}). SV, AE and MA acknowledge support from the "La Caixa" Foundation (ID~100010434, codes LCF/BQ/PI23/11970031 and LCF/BQ/PI23/11970035). SV and CAP acknowledge MICIU project PID2023-149982NB-I00 and 146453NB-I00; PP acknowledges the ERC Advanced Grant ChronoGal (GA101201412); MA acknowledges MICIU/ESF+ Ramón y Cajal grant RYC2024-049939; SS acknowledges Research Foundation-Flanders (1239522N) and FNRS (40006340). DGR acknowledges MICINN grants PID2023-149439NB-C41 and PID2023-146453NB-I00
\end{acknowledgements}

\bibliographystyle{aa} 
\bibliography{references}

@ARTICLE{2020Cretignier,
       author = {{Cretignier}, M. and {Francfort}, J. and {Dumusque}, X. and {et al.}},
        title = "{RASSINE: Interactive tool for normalising stellar spectra. I. Description and performance of the code}",
      journal = {\aap},
         year = 2020,
        month = aug,
       volume = {640},
          eid = {A42},
        pages = {A42},
          doi = {10.1051/0004-6361/202037722},
archivePrefix = {arXiv},
       eprint = {2006.13098},
 primaryClass = {astro-ph.SR},
       adsurl = {https://ui.adsabs.harvard.edu/abs/2020A&A...640A..42C}
}

@INPROCEEDINGS{1988Cayrel,
       author = {{Cayrel}, R.},
        title = "{Data Analysis}",
    booktitle = {The Impact of Very High S/N Spectroscopy on Stellar Physics},
         year = 1988,
       editor = {{Cayrel de Strobel}, G. and {Spite}, Monique},
       series = {IAU Symposium},
       volume = {132},
        month = jan,
        pages = {345},
       adsurl = {https://ui.adsabs.harvard.edu/abs/1988IAUS..132..345C}
}

@ARTICLE{2016Dotter,
       author = {{Dotter}, Aaron},
        title = "{MESA Isochrones and Stellar Tracks (MIST) 0: Methods for the Construction of Stellar Isochrones}",
      journal = {\apjs},
         year = 2016,
        month = jan,
       volume = {222},
       number = {1},
          eid = {8},
        pages = {8},
          doi = {10.3847/0067-0049/222/1/8},
archivePrefix = {arXiv},
       eprint = {1601.05144},
 primaryClass = {astro-ph.SR},
       adsurl = {https://ui.adsabs.harvard.edu/abs/2016ApJS..222....8D}
}

@ARTICLE{2016Choi,
       author = {{Choi}, Jieun and {Dotter}, Aaron and {Conroy}, Charlie and {Cantiello}, Matteo and {Paxton}, Bill and {Johnson}, Benjamin D.},
        title = "{Mesa Isochrones and Stellar Tracks (MIST). I. Solar-scaled Models}",
      journal = {\apj},
         year = 2016,
        month = jun,
       volume = {823},
       number = {2},
          eid = {102},
        pages = {102},
          doi = {10.3847/0004-637X/823/2/102},
archivePrefix = {arXiv},
       eprint = {1604.08592},
 primaryClass = {astro-ph.SR},
       adsurl = {https://ui.adsabs.harvard.edu/abs/2016ApJ...823..102C}
}

@ARTICLE{2010Pancino,
       author = {{Pancino}, E. and {Carrera}, R. and {Rossetti}, E. and {Gallart}, C.},
        title = "{Chemical abundance analysis of the open clusters Cr 110, NGC 2099 (M 37), NGC 2420, NGC 7789, and M 67 (NGC 2682)}",
      journal = {\aap},
         year = 2010,
        month = feb,
       volume = {511},
          eid = {A56},
        pages = {A56},
          doi = {10.1051/0004-6361/200912965},
archivePrefix = {arXiv},
       eprint = {0910.0723},
 primaryClass = {astro-ph.GA},
       adsurl = {https://ui.adsabs.harvard.edu/abs/2010A&A...511A..56P}
}

@ARTICLE{2017Jofre,
       author = {{Jofr{\'e}}, P. and {Heiter}, U. and {Worley}, C.~C. and {Blanco-Cuaresma}, S. and {Soubiran}, C. and {Masseron}, T. and {Hawkins}, K. and {Adibekyan}, V. and {Buder}, S. and {Casamiquela}, L. and {Gilmore}, G. and {Hourihane}, A. and {Tabernero}, H.},
        title = "{Gaia FGK benchmark stars: opening the black box of stellar element abundance determination}",
      journal = {\aap},
         year = 2017,
        month = may,
       volume = {601},
          eid = {A38},
        pages = {A38},
          doi = {10.1051/0004-6361/201629833},
archivePrefix = {arXiv},
       eprint = {1612.05013},
 primaryClass = {astro-ph.SR},
       adsurl = {https://ui.adsabs.harvard.edu/abs/2017A&A...601A..38J}
}

@ARTICLE{2008Gustafsson,
       author = {{Gustafsson}, B. and {Edvardsson}, B. and {Eriksson}, K. and {J{\o}rgensen}, U.~G. and {Nordlund}, {\r{A}}. and {Plez}, B.},
        title = "{A grid of MARCS model atmospheres for late-type stars. I. Methods and general properties}",
      journal = {\aap},
         year = 2008,
        month = aug,
       volume = {486},
       number = {3},
        pages = {951-970},
          doi = {10.1051/0004-6361:200809724},
archivePrefix = {arXiv},
       eprint = {0805.0554},
 primaryClass = {astro-ph},
       adsurl = {https://ui.adsabs.harvard.edu/abs/2008A&A...486..951G}
}

@ARTICLE{2020Santos,
       author = {{Santos-Peral}, P. and {Recio-Blanco}, A. and {de Laverny}, P. and {Fern{\'a}ndez-Alvar}, E. and {Ordenovic}, C.},
        title = "{The AMBRE Project: Spectrum normalisation influence on Mg abundances in the metal-rich Galactic disc}",
      journal = {\aap},
         year = 2020,
        month = jul,
       volume = {639},
          eid = {A140},
        pages = {A140},
          doi = {10.1051/0004-6361/202037522},
archivePrefix = {arXiv},
       eprint = {2006.08544},
 primaryClass = {astro-ph.GA},
       adsurl = {https://ui.adsabs.harvard.edu/abs/2020A&A...639A.140S}
}

@ARTICLE{2023Recio,
       author = {{Recio-Blanco}, A. and {de Laverny}, P. and {Palicio}, P.~A. and {Kordopatis}, G. and {{\'A}lvarez}, M.~A. and {Schultheis}, M. and {Contursi}, G. and {Zhao}, H. and {Torralba Elipe}, G. and {Ordenovic}, C. and {Manteiga}, M. and {Dafonte}, C. and {Oreshina-Slezak}, I. and {Bijaoui}, A. and {Fr{\'e}mat}, Y. and {Seabroke}, G. and {Pailler}, F. and {Spitoni}, E. and {Poggio}, E. and {Creevey}, O.~L. and {Abreu Aramburu}, A. and {Accart}, S. and {Andrae}, R. and {Bailer-Jones}, C.~A.~L. and {Bellas-Velidis}, I. and {Brouillet}, N. and {Brugaletta}, E. and {Burlacu}, A. and {Carballo}, R. and {Casamiquela}, L. and {Chiavassa}, A. and {Cooper}, W.~J. and {Dapergolas}, A. and {Delchambre}, L. and {Dharmawardena}, T.~E. and {Drimmel}, R. and {Edvardsson}, B. and {Fouesneau}, M. and {Garabato}, D. and {Garc{\'\i}a-Lario}, P. and {Garc{\'\i}a-Torres}, M. and {Gavel}, A. and {Gomez}, A. and {Gonz{\'a}lez-Santamar{\'\i}a}, I. and {Hatzidimitriou}, D. and {Heiter}, U. and {Jean-Antoine Piccolo}, A. and {Kontizas}, M. and {Korn}, A.~J. and {Lanzafame}, A.~C. and {Lebreton}, Y. and {Le Fustec}, Y. and {Licata}, E.~L. and {Lindstr{\o}m}, H.~E.~P. and {Livanou}, E. and {Lobel}, A. and {Lorca}, A. and {Magdaleno Romeo}, A. and {Marocco}, F. and {Marshall}, D.~J. and {Mary}, N. and {Nicolas}, C. and {Pallas-Quintela}, L. and {Panem}, C. and {Pichon}, B. and {Riclet}, F. and {Robin}, C. and {Rybizki}, J. and {Santove{\~n}a}, R. and {Silvelo}, A. and {Smart}, R.~L. and {Sarro}, L.~M. and {Sordo}, R. and {Soubiran}, C. and {S{\"u}veges}, M. and {Ulla}, A. and {Vallenari}, A. and {Zorec}, J. and {Utrilla}, E. and {Bakker}, J.},
        title = "{Gaia Data Release 3. Analysis of RVS spectra using the General Stellar Parametriser from spectroscopy}",
      journal = {\aap},
         year = 2023,
        month = jun,
       volume = {674},
          eid = {A29},
        pages = {A29},
          doi = {10.1051/0004-6361/202243750},
archivePrefix = {arXiv},
       eprint = {2206.05541},
 primaryClass = {astro-ph.GA},
       adsurl = {https://ui.adsabs.harvard.edu/abs/2023A&A...674A..29R}
}

@ARTICLE{1998Alvarez,
       author = {{Alvarez}, R. and {Plez}, B.},
        title = "{Near-infrared narrow-band photometry of M-giant and Mira stars: models meet observations}",
      journal = {\aap},
         year = 1998,
        month = feb,
       volume = {330},
        pages = {1109-1119},
          doi = {10.48550/arXiv.astro-ph/9710157},
archivePrefix = {arXiv},
       eprint = {astro-ph/9710157},
 primaryClass = {astro-ph},
       adsurl = {https://ui.adsabs.harvard.edu/abs/1998A&A...330.1109A}
}

@ARTICLE{2023Contursi,
       author = {{Contursi}, G. and {de Laverny}, P. and {Recio-Blanco}, A. and {Spitoni}, E. and {Palicio}, P.~A. and {Poggio}, E. and {Grisoni}, V. and {Cescutti}, G. and {Matteucci}, F. and {Spina}, L. and {{\'A}lvarez}, M.~A. and {Kordopatis}, G. and {Ordenovic}, C. and {Oreshina-Slezak}, I. and {Zhao}, H.},
        title = "{The cerium content of the Milky Way as revealed by Gaia DR3 GSP-Spec abundances}",
      journal = {\aap},
         year = 2023,
        month = feb,
       volume = {670},
          eid = {A106},
        pages = {A106},
          doi = {10.1051/0004-6361/202244469},
archivePrefix = {arXiv},
       eprint = {2207.05368},
 primaryClass = {astro-ph.GA},
       adsurl = {https://ui.adsabs.harvard.edu/abs/2023A&A...670A.106C}
}

@MISC{2012Plez,
       author = {{Plez}, B.},
        title = "{Turbospectrum: Code for spectral synthesis}",
 howpublished = {Astrophysics Source Code Library, record ascl:1205.004},
         year = 2012,
        month = may,
          eid = {ascl:1205.004},
        pages = {ascl:1205.004},
archivePrefix = {ascl},
       eprint = {1205.004},
       adsurl = {https://ui.adsabs.harvard.edu/abs/2012ascl.soft05004P}
}

@ARTICLE{2008Takeda,
       author = {{Takeda}, Yoichi and {Sato}, Bun'ei and {Murata}, Daisuke},
        title = "{Stellar Parameters and Elemental Abundances of Late-G Giants}",
      journal = {\pasj},
         year = 2008,
        month = aug,
       volume = {60},
        pages = {781},
          doi = {10.1093/pasj/60.4.781},
archivePrefix = {arXiv},
       eprint = {0805.2434},
 primaryClass = {astro-ph},
       adsurl = {https://ui.adsabs.harvard.edu/abs/2008PASJ...60..781T}
}

@ARTICLE{2005Takeda,
       author = {{Takeda}, Yoichi and {Honda}, Satoshi},
        title = "{Photospheric CNO Abundances of Solar-Type Stars}",
      journal = {\pasj},
         year = 2005,
        month = feb,
       volume = {57},
        pages = {65-82},
          doi = {10.1093/pasj/57.1.65},
       adsurl = {https://ui.adsabs.harvard.edu/abs/2005PASJ...57...65T}
}

@ARTICLE{2017Ting,
       author = {{Ting}, Yuan-Sen and {Conroy}, Charlie and {Rix}, Hans-Walter and {Cargile}, Phillip},
        title = "{Prospects for Measuring Abundances of >20 Elements with Low-resolution Stellar Spectra}",
      journal = {\apj},
         year = 2017,
        month = jul,
       volume = {843},
       number = {1},
          eid = {32},
        pages = {32},
          doi = {10.3847/1538-4357/aa7688},
archivePrefix = {arXiv},
       eprint = {1706.00111},
 primaryClass = {astro-ph.SR},
       adsurl = {https://ui.adsabs.harvard.edu/abs/2017ApJ...843...32T}
}

@book{Cramer1945,
  author    = {Cram{\'e}r, Harald},
  title     = {Mathematical Methods of Statistics},
  publisher = {Princeton University Press},
  address   = {Princeton, NJ},
  year      = {1945}
}

@article{Rao1945,
  author  = {Rao, C. R.},
  title   = {},
  journal = {Bulletin of the Calcutta Mathematical Society},
  volume  = {37},
  pages   = {81--91},
  year    = {1945}
}

@ARTICLE{2007Grevesse,
       author = {{Grevesse}, N. and {Asplund}, M. and {Sauval}, A.~J.},
        title = "{The Solar Chemical Composition}",
      journal = {\ssr},
         year = 2007,
        month = jun,
       volume = {130},
       number = {1-4},
        pages = {105-114},
          doi = {10.1007/s11214-007-9173-7},
       adsurl = {https://ui.adsabs.harvard.edu/abs/2007SSRv..130..105G}
}

@ARTICLE{2021Heiter,
       author = {{Heiter}, U. and {Lind}, K. and {Bergemann}, M. and {Asplund}, M. and {Mikolaitis}, {\v{S}}. and {Barklem}, P.~S. and {Masseron}, T. and {de Laverny}, P. and {Magrini}, L. and {Edvardsson}, B. and {J{\"o}nsson}, H. and {Pickering}, J.~C. and {Ryde}, N. and {Bayo Ar{\'a}n}, A. and {Bensby}, T. and {Casey}, A.~R. and {Feltzing}, S. and {Jofr{\'e}}, P. and {Korn}, A.~J. and {Pancino}, E. and {Damiani}, F. and {Lanzafame}, A. and {Lardo}, C. and {Monaco}, L. and {Morbidelli}, L. and {Smiljanic}, R. and {Worley}, C. and {Zaggia}, S. and {Randich}, S. and {Gilmore}, G.~F.},
        title = "{Atomic data for the Gaia-ESO Survey}",
      journal = {\aap},
         year = 2021,
        month = jan,
       volume = {645},
          eid = {A106},
        pages = {A106},
          doi = {10.1051/0004-6361/201936291},
archivePrefix = {arXiv},
       eprint = {2011.02049},
 primaryClass = {astro-ph.IM},
       adsurl = {https://ui.adsabs.harvard.edu/abs/2021A&A...645A.106H}
}

@ARTICLE{2016decastro,
       author = {{de Castro}, D.~B. and {Pereira}, C.~B. and {Roig}, F. and {Jilinski}, E. and {Drake}, N.~A. and {Chavero}, C. and {Sales Silva}, J.~V.},
        title = "{Chemical abundances and kinematics of barium stars}",
      journal = {\mnras},
         year = 2016,
        month = jul,
       volume = {459},
       number = {4},
        pages = {4299-4324},
          doi = {10.1093/mnras/stw815},
archivePrefix = {arXiv},
       eprint = {1604.03031},
 primaryClass = {astro-ph.SR},
       adsurl = {https://ui.adsabs.harvard.edu/abs/2016MNRAS.459.4299D}
}

@ARTICLE{2023Gaiacoll,
       author = {{Gaia Collaboration} and {Vallenari}, A. and {Brown}, A.~G.~A. and {Prusti}, T. and {de Bruijne}, J.~H.~J. and {Arenou}, F. and {Babusiaux}, C. and {Biermann}, M. and {Creevey}, O.~L. and {Ducourant}, C. and {Evans}, D.~W. and {Eyer}, L. and {Guerra}, R. and {Hutton}, A. and {Jordi}, C. and {Klioner}, S.~A. and {Lammers}, U.~L. and {Lindegren}, L. and {Luri}, X. and {Mignard}, F. and {Panem}, C. and {Pourbaix}, D. and {Randich}, S. and {Sartoretti}, P. and {Soubiran}, C. and {Tanga}, P. and {Walton}, N.~A. and {Bailer-Jones}, C.~A.~L. and {Bastian}, U. and {Drimmel}, R. and {Jansen}, F. and {Katz}, D. and {Lattanzi}, M.~G. and {van Leeuwen}, F. and {Bakker}, J. and {Cacciari}, C. and {Casta{\~n}eda}, J. and {De Angeli}, F. and {Fabricius}, C. and {Fouesneau}, M. and {Fr{\'e}mat}, Y. and {Galluccio}, L. and {Guerrier}, A. and {Heiter}, U. and {Masana}, E. and {Messineo}, R. and {Mowlavi}, N. and {Nicolas}, C. and {Nienartowicz}, K. and {Pailler}, F. and {Panuzzo}, P. and {Riclet}, F. and {Roux}, W. and {Seabroke}, G.~M. and {Sordo}, R. and {Th{\'e}venin}, F. and {Gracia-Abril}, G. and {Portell}, J. and {Teyssier}, D. and {Altmann}, M. and {Andrae}, R. and {Audard}, M. and {Bellas-Velidis}, I. and {Benson}, K. and {Berthier}, J. and {Blomme}, R. and {Burgess}, P.~W. and {Busonero}, D. and {Busso}, G. and {C{\'a}novas}, H. and {Carry}, B. and {Cellino}, A. and {Cheek}, N. and {Clementini}, G. and {Damerdji}, Y. and {Davidson}, M. and {de Teodoro}, P. and {Nu{\~n}ez Campos}, M. and {Delchambre}, L. and {Dell'Oro}, A. and {Esquej}, P. and {Fern{\'a}ndez-Hern{\'a}ndez}, J. and {Fraile}, E. and {Garabato}, D. and {Garc{\'\i}a-Lario}, P. and {Gosset}, E. and {Haigron}, R. and {Halbwachs}, J.-L. and {Hambly}, N.~C. and {Harrison}, D.~L. and {Hern{\'a}ndez}, J. and {Hestroffer}, D. and {Hodgkin}, S.~T. and {Holl}, B. and {Jan{\ss}en}, K. and {Jevardat de Fombelle}, G. and {Jordan}, S. and {Krone-Martins}, A. and {Lanzafame}, A.~C. and {L{\"o}ffler}, W. and {Marchal}, O. and {Marrese}, P.~M. and {Moitinho}, A. and {Muinonen}, K. and {Osborne}, P. and {Pancino}, E. and {Pauwels}, T. and {Recio-Blanco}, A. and {Reyl{\'e}}, C. and {Riello}, M. and {Rimoldini}, L. and {Roegiers}, T. and {Rybizki}, J. and {Sarro}, L.~M. and {Siopis}, C. and {Smith}, M. and {Sozzetti}, A. and {Utrilla}, E. and {van Leeuwen}, M. and {Abbas}, U. and {{\'A}brah{\'a}m}, P. and {Abreu Aramburu}, A. and {Aerts}, C. and {Aguado}, J.~J. and {Ajaj}, M. and {Aldea-Montero}, F. and {Altavilla}, G. and {{\'A}lvarez}, M.~A. and {Alves}, J. and {Anders}, F. and {Anderson}, R.~I. and {Anglada Varela}, E. and {Antoja}, T. and {Baines}, D. and {Baker}, S.~G. and {Balaguer-N{\'u}{\~n}ez}, L. and {Balbinot}, E. and {Balog}, Z. and {Barache}, C. and {Barbato}, D. and {Barros}, M. and {Barstow}, M.~A. and {Bartolom{\'e}}, S. and {Bassilana}, J.-L. and {Bauchet}, N. and {Becciani}, U. and {Bellazzini}, M. and {Berihuete}, A. and {Bernet}, M. and {Bertone}, S. and {Bianchi}, L. and {Binnenfeld}, A. and {Blanco-Cuaresma}, S. and {Blazere}, A. and {Boch}, T. and {Bombrun}, A. and {Bossini}, D. and {Bouquillon}, S. and {Bragaglia}, A. and {Bramante}, L. and {Breedt}, E. and {Bressan}, A. and {Brouillet}, N. and {Brugaletta}, E. and {Bucciarelli}, B. and {Burlacu}, A. and {Butkevich}, A.~G. and {Buzzi}, R. and {Caffau}, E. and {Cancelliere}, R. and {Cantat-Gaudin}, T. and {Carballo}, R. and {Carlucci}, T. and {Carnerero}, M.~I. and {Carrasco}, J.~M. and {Casamiquela}, L. and {Castellani}, M. and {Castro-Ginard}, A. and {Chaoul}, L. and {Charlot}, P. and {Chemin}, L. and {Chiaramida}, V. and {Chiavassa}, A. and {Chornay}, N. and {Comoretto}, G. and {Contursi}, G. and {Cooper}, W.~J. and {Cornez}, T. and {Cowell}, S. and {Crifo}, F. and {Cropper}, M. and {Crosta}, M. and {Crowley}, C. and {Dafonte}, C. and {Dapergolas}, A. and {David}, M. and {David}, P. and {de Laverny}, P. and {De Luise}, F. and {De March}, R.},
        title = "{Gaia Data Release 3. Summary of the content and survey properties}",
      journal = {\aap},
         year = 2023,
        month = jun,
       volume = {674},
          eid = {A1},
        pages = {A1},
          doi = {10.1051/0004-6361/202243940},
archivePrefix = {arXiv},
       eprint = {2208.00211},
 primaryClass = {astro-ph.GA},
       adsurl = {https://ui.adsabs.harvard.edu/abs/2023A&A...674A...1G}
}

@ARTICLE{2023Lucey,
       author = {{Lucey}, Madeline and {Al Kharusi}, Nariman and {Hawkins}, Keith and {Ting}, Yuan-Sen and {Ramachandra}, Nesar and {Price-Whelan}, Adrian M. and {Beers}, Timothy C. and {Lee}, Young Sun and {Yoon}, Jinmi},
        title = "{Carbon-enhanced metal-poor star candidates from BP/RP spectra in Gaia DR3}",
      journal = {\mnras},
         year = 2023,
        month = aug,
       volume = {523},
       number = {3},
        pages = {4049-4066},
          doi = {10.1093/mnras/stad1675},
archivePrefix = {arXiv},
       eprint = {2206.08299},
 primaryClass = {astro-ph.GA},
       adsurl = {https://ui.adsabs.harvard.edu/abs/2023MNRAS.523.4049L}
}

@ARTICLE{1980McClure,
       author = {{McClure}, R.~D. and {Fletcher}, J.~M. and {Nemec}, J.~M.},
        title = "{The binary nature of the barium stars.}",
      journal = {\apjl},
         year = 1980,
        month = may,
       volume = {238},
        pages = {L35-L38},
          doi = {10.1086/183252},
       adsurl = {https://ui.adsabs.harvard.edu/abs/1980ApJ...238L..35M}
}

@ARTICLE{2016Gaia,
       author = {{Gaia Collaboration} and {Prusti}, T. and {de Bruijne}, J.~H.~J. and {Brown}, A.~G.~A. and {Vallenari}, A. and {Babusiaux}, C. and {Bailer-Jones}, C.~A.~L. and {Bastian}, U. and {Biermann}, M. and {Evans}, D.~W. and {Eyer}, L. and {Jansen}, F. and {Jordi}, C. and {Klioner}, S.~A. and {Lammers}, U. and {Lindegren}, L. and {Luri}, X. and {Mignard}, F. and {Milligan}, D.~J. and {Panem}, C. and {Poinsignon}, V. and {Pourbaix}, D. and {Randich}, S. and {Sarri}, G. and {Sartoretti}, P. and {Siddiqui}, H.~I. and {Soubiran}, C. and {Valette}, V. and {van Leeuwen}, F. and {Walton}, N.~A. and {Aerts}, C. and {Arenou}, F. and {Cropper}, M. and {Drimmel}, R. and {H{\o}g}, E. and {Katz}, D. and {Lattanzi}, M.~G. and {O'Mullane}, W. and {Grebel}, E.~K. and {Holland}, A.~D. and {Huc}, C. and {Passot}, X. and {Bramante}, L. and {Cacciari}, C. and {Casta{\~n}eda}, J. and {Chaoul}, L. and {Cheek}, N. and {De Angeli}, F. and {Fabricius}, C. and {Guerra}, R. and {Hern{\'a}ndez}, J. and {Jean-Antoine-Piccolo}, A. and {Masana}, E. and {Messineo}, R. and {Mowlavi}, N. and {Nienartowicz}, K. and {Ord{\'o}{\~n}ez-Blanco}, D. and {Panuzzo}, P. and {Portell}, J. and {Richards}, P.~J. and {Riello}, M. and {Seabroke}, G.~M. and {Tanga}, P. and {Th{\'e}venin}, F. and {Torra}, J. and {Els}, S.~G. and {Gracia-Abril}, G. and {Comoretto}, G. and {Garcia-Reinaldos}, M. and {Lock}, T. and {Mercier}, E. and {Altmann}, M. and {Andrae}, R. and {Astraatmadja}, T.~L. and {Bellas-Velidis}, I. and {Benson}, K. and {Berthier}, J. and {Blomme}, R. and {Busso}, G. and {Carry}, B. and {Cellino}, A. and {Clementini}, G. and {Cowell}, S. and {Creevey}, O. and {Cuypers}, J. and {Davidson}, M. and {De Ridder}, J. and {de Torres}, A. and {Delchambre}, L. and {Dell'Oro}, A. and {Ducourant}, C. and {Fr{\'e}mat}, Y. and {Garc{\'\i}a-Torres}, M. and {Gosset}, E. and {Halbwachs}, J.-L. and {Hambly}, N.~C. and {Harrison}, D.~L. and {Hauser}, M. and {Hestroffer}, D. and {Hodgkin}, S.~T. and {Huckle}, H.~E. and {Hutton}, A. and {Jasniewicz}, G. and {Jordan}, S. and {Kontizas}, M. and {Korn}, A.~J. and {Lanzafame}, A.~C. and {Manteiga}, M. and {Moitinho}, A. and {Muinonen}, K. and {Osinde}, J. and {Pancino}, E. and {Pauwels}, T. and {Petit}, J.-M. and {Recio-Blanco}, A. and {Robin}, A.~C. and {Sarro}, L.~M. and {Siopis}, C. and {Smith}, M. and {Smith}, K.~W. and {Sozzetti}, A. and {Thuillot}, W. and {van Reeven}, W. and {Viala}, Y. and {Abbas}, U. and {Abreu Aramburu}, A. and {Accart}, S. and {Aguado}, J.~J. and {Allan}, P.~M. and {Allasia}, W. and {Altavilla}, G. and {{\'A}lvarez}, M.~A. and {Alves}, J. and {Anderson}, R.~I. and {Andrei}, A.~H. and {Anglada Varela}, E. and {Antiche}, E. and {Antoja}, T. and {Ant{\'o}n}, S. and {Arcay}, B. and {Atzei}, A. and {Ayache}, L. and {Bach}, N. and {Baker}, S.~G. and {Balaguer-N{\'u}{\~n}ez}, L. and {Barache}, C. and {Barata}, C. and {Barbier}, A. and {Barblan}, F. and {Baroni}, M. and {Barrado y Navascu{\'e}s}, D. and {Barros}, M. and {Barstow}, M.~A. and {Becciani}, U. and {Bellazzini}, M. and {Bellei}, G. and {Bello Garc{\'\i}a}, A. and {Belokurov}, V. and {Bendjoya}, P. and {Berihuete}, A. and {Bianchi}, L. and {Bienaym{\'e}}, O. and {Billebaud}, F. and {Blagorodnova}, N. and {Blanco-Cuaresma}, S. and {Boch}, T. and {Bombrun}, A. and {Borrachero}, R. and {Bouquillon}, S. and {Bourda}, G. and {Bouy}, H. and {Bragaglia}, A. and {Breddels}, M.~A. and {Brouillet}, N. and {Br{\"u}semeister}, T. and {Bucciarelli}, B. and {Budnik}, F. and {Burgess}, P. and {Burgon}, R. and {Burlacu}, A. and {Busonero}, D. and {Buzzi}, R. and {Caffau}, E. and {Cambras}, J. and {Campbell}, H. and {Cancelliere}, R. and {Cantat-Gaudin}, T. and {Carlucci}, T. and {Carrasco}, J.~M. and {Castellani}, M. and {Charlot}, P. and {Charnas}, J. and {Charvet}, P. and {Chassat}, F. and {Chiavassa}, A. and {Clotet}, M. and {Cocozza}, G. and {Collins}, R.~S. and {Collins}, P. and {Costigan}, G.},
        title = "{The Gaia mission}",
      journal = {\aap},
         year = 2016,
        month = nov,
       volume = {595},
          eid = {A1},
        pages = {A1},
          doi = {10.1051/0004-6361/201629272},
archivePrefix = {arXiv},
       eprint = {1609.04153},
 primaryClass = {astro-ph.IM},
       adsurl = {https://ui.adsabs.harvard.edu/abs/2016A&A...595A...1G}
}

@ARTICLE{2026Levine,
       author = {{Levine}, Jaden and {Manea}, Catherine and {Hawkins}, Keith and {Sullivan}, Kendall and {Rubin}, Kate H.~R. and {Maas}, Zachary and {Nine}, Andrew C.},
        title = "{Barium Stars across the Milky Way: Probing Their Origins via the GALAH Survey}",
      journal = {\apj},
         year = 2026,
        month = mar,
       volume = {1000},
       number = {1},
          eid = {117},
        pages = {117},
          doi = {10.3847/1538-4357/ae4726},
archivePrefix = {arXiv},
       eprint = {2602.15107},
 primaryClass = {astro-ph.GA},
       adsurl = {https://ui.adsabs.harvard.edu/abs/2026ApJ..1000..117L}
}

@ARTICLE{2025Arentsen,
       author = {{Ardern-Arentsen}, Anke and {Kane}, Sarah G. and {Belokurov}, Vasily and {Matsuno}, Tadafumi and {Montelius}, Martin and {Monty}, Stephanie and {Sanders}, Jason L.},
        title = "{Predicting metallicities and carbon abundances from Gaia XP spectra for (carbon-enhanced) metal-poor stars}",
      journal = {\mnras},
         year = 2025,
        month = feb,
       volume = {537},
       number = {2},
        pages = {1984-2002},
          doi = {10.1093/mnras/staf096},
archivePrefix = {arXiv},
       eprint = {2410.11077},
 primaryClass = {astro-ph.SR},
       adsurl = {https://ui.adsabs.harvard.edu/abs/2025MNRAS.537.1984A}
}

@ARTICLE{2023Katz,
       author = {{Katz}, D. and {Sartoretti}, P. and {Guerrier}, A. and {Panuzzo}, P. and {Seabroke}, G.~M. and {Th{\'e}venin}, F. and {Cropper}, M. and {Benson}, K. and {Blomme}, R. and {Haigron}, R. and {Marchal}, O. and {Smith}, M. and {Baker}, S. and {Chemin}, L. and {Damerdji}, Y. and {David}, M. and {Dolding}, C. and {Fr{\'e}mat}, Y. and {Gosset}, E. and {Jan{\ss}en}, K. and {Jasniewicz}, G. and {Lobel}, A. and {Plum}, G. and {Samaras}, N. and {Snaith}, O. and {Soubiran}, C. and {Vanel}, O. and {Zwitter}, T. and {Antoja}, T. and {Arenou}, F. and {Babusiaux}, C. and {Brouillet}, N. and {Caffau}, E. and {Di Matteo}, P. and {Fabre}, C. and {Fabricius}, C. and {Fragkoudi}, F. and {Haywood}, M. and {Huckle}, H.~E. and {Hottier}, C. and {Lasne}, Y. and {Leclerc}, N. and {Mastrobuono-Battisti}, A. and {Royer}, F. and {Teyssier}, D. and {Zorec}, J. and {Crifo}, F. and {Jean-Antoine Piccolo}, A. and {Turon}, C. and {Viala}, Y.},
        title = "{Gaia Data Release 3. Properties and validation of the radial velocities}",
      journal = {\aap},
         year = 2023,
        month = jun,
       volume = {674},
          eid = {A5},
        pages = {A5},
          doi = {10.1051/0004-6361/202244220},
archivePrefix = {arXiv},
       eprint = {2206.05902},
 primaryClass = {astro-ph.GA},
       adsurl = {https://ui.adsabs.harvard.edu/abs/2023A&A...674A...5K}
}

@ARTICLE{2018Cropper,
       author = {{Cropper}, M. and {Katz}, D. and {Sartoretti}, P. and {Prusti}, T. and {de Bruijne}, J.~H.~J. and {Chassat}, F. and {Charvet}, P. and {Boyadjian}, J. and {Perryman}, M. and {Sarri}, G. and {Gare}, P. and {Erdmann}, M. and {Munari}, U. and {Zwitter}, T. and {Wilkinson}, M. and {Arenou}, F. and {Vallenari}, A. and {G{\'o}mez}, A. and {Panuzzo}, P. and {Seabroke}, G. and {Allende Prieto}, C. and {Benson}, K. and {Marchal}, O. and {Huckle}, H. and {Smith}, M. and {Dolding}, C. and {Jan{\ss}en}, K. and {Viala}, Y. and {Blomme}, R. and {Baker}, S. and {Boudreault}, S. and {Crifo}, F. and {Soubiran}, C. and {Fr{\'e}mat}, Y. and {Jasniewicz}, G. and {Guerrier}, A. and {Guy}, L.~P. and {Turon}, C. and {Jean-Antoine-Piccolo}, A. and {Th{\'e}venin}, F. and {David}, M. and {Gosset}, E. and {Damerdji}, Y.},
        title = "{Gaia Data Release 2. Gaia Radial Velocity Spectrometer}",
      journal = {\aap},
         year = 2018,
        month = aug,
       volume = {616},
          eid = {A5},
        pages = {A5},
          doi = {10.1051/0004-6361/201832763},
archivePrefix = {arXiv},
       eprint = {1804.09369},
 primaryClass = {astro-ph.IM},
       adsurl = {https://ui.adsabs.harvard.edu/abs/2018A&A...616A...5C}
}

@ARTICLE{2023Creevey,
       author = {{Gaia Collaboration} and {Creevey}, O.~L. and {Sarro}, L.~M. and {Lobel}, A. and {Pancino}, E. and {Andrae}, R. and {Smart}, R.~L. and {Clementini}, G. and {Heiter}, U. and {Korn}, A.~J. and {Fouesneau}, M. and {Fr{\'e}mat}, Y. and {De Angeli}, F. and {Vallenari}, A. and {Harrison}, D.~L. and {Th{\'e}venin}, F. and {Reyl{\'e}}, C. and {Sordo}, R. and {Garofalo}, A. and {Brown}, A.~G.~A. and {Eyer}, L. and {Prusti}, T. and {de Bruijne}, J.~H.~J. and {Arenou}, F. and {Babusiaux}, C. and {Biermann}, M. and {Ducourant}, C. and {Evans}, D.~W. and {Guerra}, R. and {Hutton}, A. and {Jordi}, C. and {Klioner}, S.~A. and {Lammers}, U.~L. and {Lindegren}, L. and {Luri}, X. and {Mignard}, F. and {Panem}, C. and {Pourbaix}, D. and {Randich}, S. and {Sartoretti}, P. and {Soubiran}, C. and {Tanga}, P. and {Walton}, N.~A. and {Bailer-Jones}, C.~A.~L. and {Bastian}, U. and {Drimmel}, R. and {Jansen}, F. and {Katz}, D. and {Lattanzi}, M.~G. and {van Leeuwen}, F. and {Bakker}, J. and {Cacciari}, C. and {Casta{\~n}eda}, J. and {Fabricius}, C. and {Galluccio}, L. and {Guerrier}, A. and {Masana}, E. and {Messineo}, R. and {Mowlavi}, N. and {Nicolas}, C. and {Nienartowicz}, K. and {Pailler}, F. and {Panuzzo}, P. and {Riclet}, F. and {Roux}, W. and {Seabroke}, G.~M. and {Gracia-Abril}, G. and {Portell}, J. and {Teyssier}, D. and {Altmann}, M. and {Audard}, M. and {Bellas-Velidis}, I. and {Benson}, K. and {Berthier}, J. and {Blomme}, R. and {Burgess}, P.~W. and {Busonero}, D. and {Busso}, G. and {C{\'a}novas}, H. and {Carry}, B. and {Cellino}, A. and {Cheek}, N. and {Damerdji}, Y. and {Davidson}, M. and {de Teodoro}, P. and {Nu{\~n}ez Campos}, M. and {Delchambre}, L. and {Dell'Oro}, A. and {Esquej}, P. and {Fern{\'a}ndez-Hern{\'a}ndez}, J. and {Fraile}, E. and {Garabato}, D. and {Garc{\'\i}a-Lario}, P. and {Gosset}, E. and {Haigron}, R. and {Halbwachs}, J.-L. and {Hambly}, N.~C. and {Hern{\'a}ndez}, J. and {Hestroffer}, D. and {Hodgkin}, S.~T. and {Holl}, B. and {Jan{\ss}en}, K. and {Jevardat de Fombelle}, G. and {Jordan}, S. and {Krone-Martins}, A. and {Lanzafame}, A.~C. and {L{\"o}ffler}, W. and {Marchal}, O. and {Marrese}, P.~M. and {Moitinho}, A. and {Muinonen}, K. and {Osborne}, P. and {Pauwels}, T. and {Recio-Blanco}, A. and {Riello}, M. and {Rimoldini}, L. and {Roegiers}, T. and {Rybizki}, J. and {Siopis}, C. and {Smith}, M. and {Sozzetti}, A. and {Utrilla}, E. and {van Leeuwen}, M. and {Abbas}, U. and {{\'A}brah{\'a}m}, P. and {Abreu Aramburu}, A. and {Aerts}, C. and {Aguado}, J.~J. and {Ajaj}, M. and {Aldea-Montero}, F. and {Altavilla}, G. and {{\'A}lvarez}, M.~A. and {Alves}, J. and {Anders}, F. and {Anderson}, R.~I. and {Anglada Varela}, E. and {Antoja}, T. and {Baines}, D. and {Baker}, S.~G. and {Balaguer-N{\'u}{\~n}ez}, L. and {Balbinot}, E. and {Balog}, Z. and {Barache}, C. and {Barbato}, D. and {Barros}, M. and {Barstow}, M.~A. and {Bartolom{\'e}}, S. and {Bassilana}, J.-L. and {Bauchet}, N. and {Becciani}, U. and {Bellazzini}, M. and {Berihuete}, A. and {Bernet}, M. and {Bertone}, S. and {Bianchi}, L. and {Binnenfeld}, A. and {Blanco-Cuaresma}, S. and {Boch}, T. and {Bombrun}, A. and {Bossini}, D. and {Bouquillon}, S. and {Bragaglia}, A. and {Bramante}, L. and {Breedt}, E. and {Bressan}, A. and {Brouillet}, N. and {Brugaletta}, E. and {Bucciarelli}, B. and {Burlacu}, A. and {Butkevich}, A.~G. and {Buzzi}, R. and {Caffau}, E. and {Cancelliere}, R. and {Cantat-Gaudin}, T. and {Carballo}, R. and {Carlucci}, T. and {Carnerero}, M.~I. and {Carrasco}, J.~M. and {Casamiquela}, L. and {Castellani}, M. and {Castro-Ginard}, A. and {Chaoul}, L. and {Charlot}, P. and {Chemin}, L. and {Chiaramida}, V. and {Chiavassa}, A. and {Chornay}, N. and {Comoretto}, G. and {Contursi}, G. and {Cooper}, W.~J. and {Cornez}, T. and {Cowell}, S. and {Crifo}, F. and {Cropper}, M. and {Crosta}, M. and {Crowley}, C. and {Dafonte}, C. and {Dapergolas}, A. and {David}, P. and {de Laverny}, P.},
        title = "{Gaia Data Release 3. A golden sample of astrophysical parameters}",
      journal = {\aap},
         year = 2023,
        month = jun,
       volume = {674},
          eid = {A39},
        pages = {A39},
          doi = {10.1051/0004-6361/202243800},
archivePrefix = {arXiv},
       eprint = {2206.05870},
 primaryClass = {astro-ph.SR},
       adsurl = {https://ui.adsabs.harvard.edu/abs/2023A&A...674A..39G}
}

@ARTICLE{2026Das,
       author = {{Das}, Pradosh Barun and {Zucker}, Daniel B. and {Mura-Guzm{\'a}n}, Aldo and {Borsato}, Nicholas W. and {de Silva}, Gayandhi M. and {Buder}, Sven and {Feuillet}, Diane and {Nordlander}, Thomas and {Ness}, Melissa K. and {Martell}, Sarah L. and {Kos}, Janez and {Bland-Hawthorn}, Joss and {Freeman}, Ken C. and {Casey}, Andrew R. and {Lewis}, Geraint F. and {Stello}, Dennis and {de Grijs}, Richard and {GALAH Collaboration}},
        title = "{The GALAH Survey: Neutron-Capture Elemental Abundances for 350 000 Gaia-RVS spectra and the Chemodynamics of Accreted Structures}",
      journal = {\mnras},
         year = 2026,
        month = aug,
       volume = {550},
       number = {4},
          eid = {stag1142},
        pages = {stag1142},
          doi = {10.1093/mnras/stag1142},
archivePrefix = {arXiv},
       eprint = {2606.04567},
 primaryClass = {astro-ph.GA},
       adsurl = {https://ui.adsabs.harvard.edu/abs/2026MNRAS.550g1142D}
}

@ARTICLE{2019Forsberg,
       author = {{Forsberg}, R. and {J{\"o}nsson}, H. and {Ryde}, N. and {Matteucci}, F.},
        title = "{Abundances of disk and bulge giants from high-resolution optical spectra. IV. Zr, La, Ce, Eu}",
      journal = {\aap},
         year = 2019,
        month = nov,
       volume = {631},
          eid = {A113},
        pages = {A113},
          doi = {10.1051/0004-6361/201936343},
archivePrefix = {arXiv},
       eprint = {1909.10535},
 primaryClass = {astro-ph.GA},
       adsurl = {https://ui.adsabs.harvard.edu/abs/2019A&A...631A.113F}
}

@ARTICLE{2011Bisterzo,
       author = {{Bisterzo}, S. and {Gallino}, R. and {Straniero}, O. and {Cristallo}, S. and {K{\"a}ppeler}, F.},
        title = "{The s-process in low-metallicity stars - II. Interpretation of high-resolution spectroscopic observations with asymptotic giant branch models}",
      journal = {\mnras},
         year = 2011,
        month = nov,
       volume = {418},
       number = {1},
        pages = {284-319},
          doi = {10.1111/j.1365-2966.2011.19484.x},
archivePrefix = {arXiv},
       eprint = {1108.0500},
 primaryClass = {astro-ph.SR},
       adsurl = {https://ui.adsabs.harvard.edu/abs/2011MNRAS.418..284B}
}

@ARTICLE{2025Navarrete,
       author = {{Navarrete}, Camila and {Recio-Blanco}, Alejandra and {de Laverny}, Patrick and {et al.}},
        title = "{A comprehensive Gaia view of ellipsoidal and rotational red giant binaries}",
      journal = {\aap},
         year = 2025,
        month = apr,
       volume = {696},
          eid = {A82},
        pages = {A82},
          doi = {10.1051/0004-6361/202453394},
archivePrefix = {arXiv},
       eprint = {2502.11667},
 primaryClass = {astro-ph.SR},
       adsurl = {https://ui.adsabs.harvard.edu/abs/2025A&A...696A..82N}
}

@ARTICLE{2003Sneden,
       author = {{Sneden}, Christopher and {Cowan}, John J.},
        title = "{Genesis of the Heaviest Elements in the Milky Way Galaxy}",
      journal = {Science},
         year = 2003,
        month = jan,
       volume = {299},
       number = {5603},
        pages = {70-75},
          doi = {10.1126/science.1077506},
       adsurl = {https://ui.adsabs.harvard.edu/abs/2003Sci...299...70S}
}

@ARTICLE{2024Contursi,
       author = {{Contursi}, G. and {de Laverny}, P. and {Recio-Blanco}, A. and {et al.}},
        title = "{Production of s-process elements in asymptotic giant branch stars as revealed by Gaia/GSP-Spec abundances}",
      journal = {\aap},
         year = 2024,
        month = mar,
       volume = {683},
          eid = {A138},
        pages = {A138},
          doi = {10.1051/0004-6361/202347188},
       adsurl = {https://ui.adsabs.harvard.edu/abs/2024A&A...683A.138C}
}

@ARTICLE{2026Escorza,
       author = {{Escorza}, A. and {Vitali}, S. and {Godoy-Rivera}, D. and {Shetye}, S. and {Van Winckel}, H. and {Bustos}, G. and {Goriely}, S. and {Siess}, L. and {Abdul-Masih}, M. and {Masseron}, T. and {Garc{\'\i}a-Hern{\'a}ndez}, D.~A. and {Ardern-Arentsen}, A. and {Jofr{\'e}}, P. and {Van Eck}, S.},
        title = "{Stellar nucleosynthesis in the era of large surveys: S-process-polluted binaries in GALAH DR4}",
      journal = {\aap},
         year = 2026,
        month = aug,
       volume = {712},
          eid = {A111},
        pages = {A111},
          doi = {10.1051/0004-6361/202660464},
archivePrefix = {arXiv},
       eprint = {2604.15197},
 primaryClass = {astro-ph.SR},
       adsurl = {https://ui.adsabs.harvard.edu/abs/2026A&A...712A.111E}
}

\begin{appendix}
\section{Synthetic grids}\label{append:synthesis}

We used Turbospectrum \citep{1998Alvarez,2012Plez}, with the MARCS stellar atmosphere grid \citep[][spherical geometry for giants and plane-parallel for dwarfs]{2008Gustafsson}, the Gaia-ESO Survey v6 atomic line list \citep{2021Heiter} including hyperfine structure and isotopic splitting, and the solar abundance scale of \citet{2007Grevesse}. The Ce atomic data are the same as in Table 1 of \cite{2023Contursi}. Our spectral synthesis assumes local thermodynamic equilibrium (LTE), consistent with the approach adopted by the GSP-Spec pipeline.

\begin{figure}[b]
    \centering
  \includegraphics[width=\columnwidth]{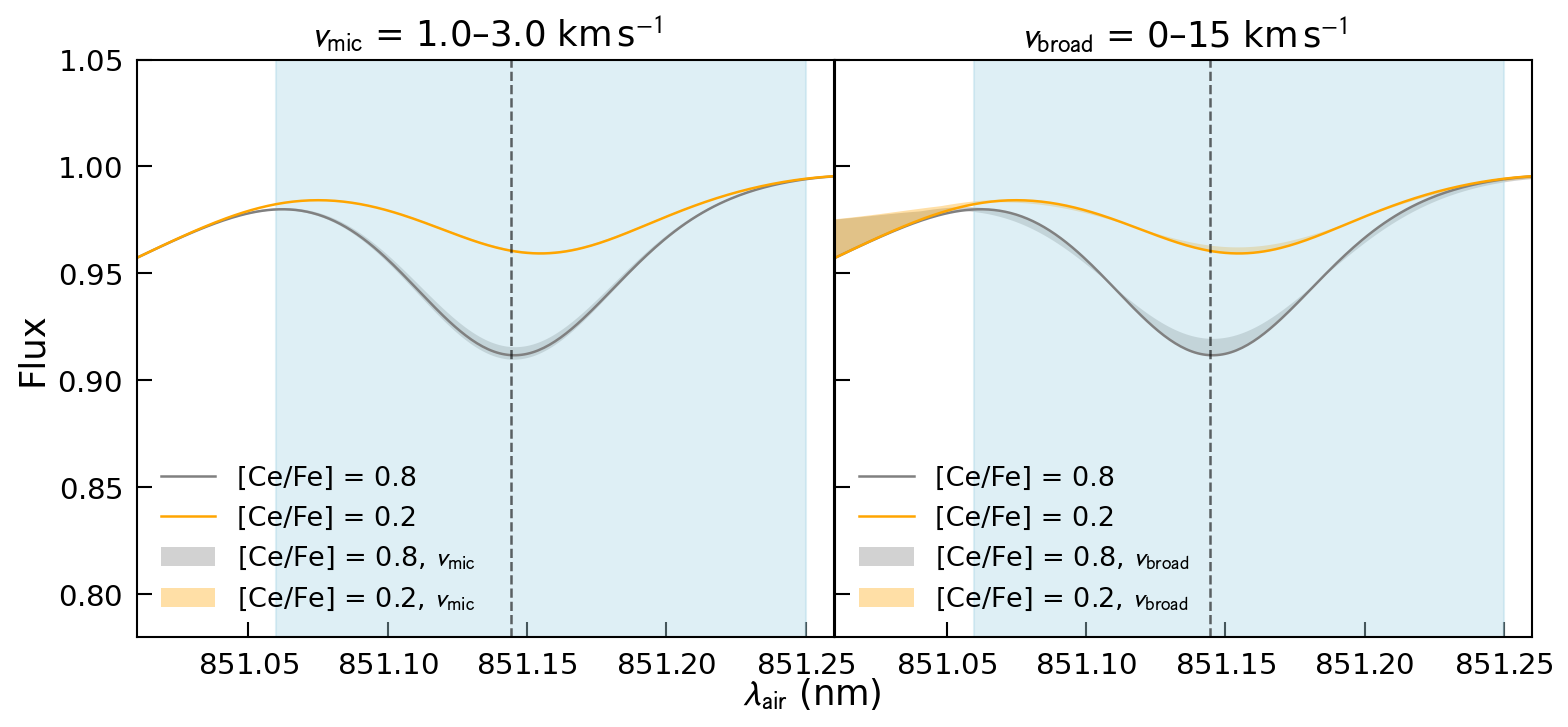}  
    \caption{Examples of two synthetic spectra of a giant star (\Teff\ = 4500~K, \logg\ = 1, [Fe/H] = $-0.2$~dex) for the two Ce abundances reported in the legend (solid lines). Shaded bands show the range of line profiles obtained when varying the microturbulence velocity ($v_{\rm mic}=1$--$3\,\mathrm{km\,s^{-1}}$, left panel) and the broadening velocity ($v_{\rm broad}=0$--$15\,\mathrm{km\,s^{-1}}$, right panel). The dashed vertical line marks the centroid of the Ce feature (an unresolved triplet of Ce~\textsc{ii} lines at 851.11~nm), and the shaded window marks the region in which the gradient is computed.}\label{fig:sp}
\end{figure}

Following RB23, we adopted the standard MARCS metallicity-dependent $\alpha$-enhancement, which increases towards lower [M/H] to reproduce the Galactic $\alpha$-enhancement trends. No rotation or macroturbulence were included. Since the empirical formula for the microturbulence velocity ($v_{\rm mic}$) adopted in RB23 is not public, we assumed values consistent with those derived in works such as \citet{2005Takeda,2008Takeda}, adopting $\mathrm{v_{\rm mic}}= 1\,\mathrm{km\,s^{-1}}$ for dwarfs and $= 2\,\mathrm{km\,s^{-1}}$ for giants. Given that Ce lines in this spectral range are generally weak, we verified by varying the broadening velocities that their effects produce a negligible change in the line profile at the spectral resolution of the RVS. Examples of synthetic spectra are illustrated in Fig.~\ref{fig:sp}.

\section{Cram\'er--Rao precision bound}\label{appendix:gradient}

Similar to \cite{2017Ting}, we impose a gradient criterion to assess whether the Ce feature carries detectable spectral information. We compute, for each node (\Teff, \logg, [Fe/H]) in the grid, the spectral gradient of the normalized flux $f_{i}$ with respect to the Ce abundance at each wavelength pixel $\lambda_{i}$ in the Ce window (851.06-851.25\,nm, see Fig.~\ref{fig:sp}):
\begin{gather}\label{eq:gradient}
    g_{i}(\rm{[Ce/Fe]}) = \partial f_i / \partial [Ce/Fe]  \cong\frac{f_{i}([\rm{Ce/Fe}]+\Delta)-f_{i}([\rm{Ce/Fe}])}{\Delta} 
\end{gather}
where $\Delta = 0.1\,$dex is the grid step. This measures the spectrum's local sensitivity to Ce variations at each grid step from [Ce/Fe] =-0.5 to +2.0 dex.

From the gradient spectrum we compute the Fisher information for the Ce
abundance label, assuming Gaussian, uncorrelated pixel noise with per-pixel
scatter $1/\mathrm{SNR}$:
\begin{equation}\label{eq:fisher}
    I(\mathrm{[Ce/Fe]}) = \mathrm{SNR}^{2} \sum_{i} g_{i}^{2}.
\end{equation}
The CRB then gives the minimum uncertainty achievable by any
unbiased estimator of [Ce/Fe],
\begin{equation}\label{eq:cramer}
    \sigma_{\mathrm{[Ce/Fe]}} \geq \frac{1}{\sqrt{I(\mathrm{[Ce/Fe]})}}
    = \frac{1}{\mathrm{SNR}\,\sqrt{\sum_{i} g_{i}^{2}}}.
\end{equation}
We consider Ce detectable at a grid when this bound falls at or below a target
precision
\begin{equation}\label{eq:detect}
    \sigma_{\mathrm{[Ce/Fe]}} \leq X\,\mathrm{dex},
\end{equation}
so that $X$, together with the assumed SNR, sets the precision required for a
meaningful abundance measurement. This defines (together with the Cayrel
detection criterion), at a given SNR, the region of atmospheric parameters over
which the RVS spectrum carries enough Ce information to recover [Ce/Fe] to better than $X$~dex and yield an individually significant feature. The bound is a theoretical floor: it assumes a locally linear response of the spectrum to
[Ce/Fe], so the actual uncertainty is equal or larger, particularly where the Ce response is nonmonotonic (Fig.~\ref{fig:gradient_bin}).

\section{Spectral response to Ce}\label{appendix:spectral}

\begin{figure*}[t]
    \centering   \includegraphics[width=\textwidth]{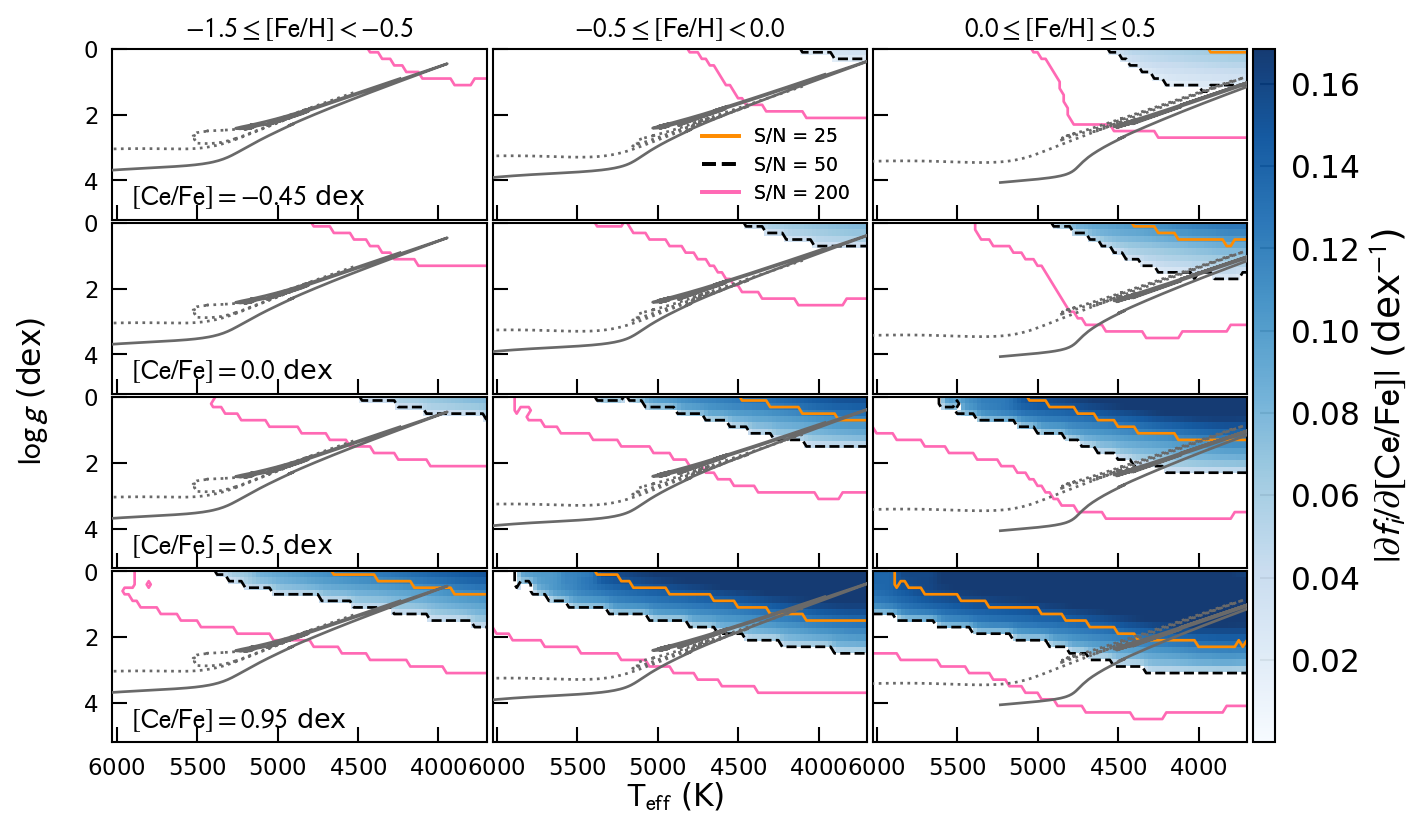}
    \caption{Similar to Fig.~\ref{fig:kiel_sensitivity}, but extended to the full range of [Ce/Fe] values spanning the synthesis grid. The color scale is the same as in Fig.~\ref{fig:kiel_sensitivity}. Solid and dotted curves show MIST evolutionary tracks \citep{2016Choi,2016Dotter} for $1\,\mathrm{M_\odot}$ and $2\,\mathrm{M_\odot}$, respectively, computed for [Fe/H] = $-1.0$, $-0.5;-0.25$, and $+0.5;+0.25$ in the left, middle, and right panels. The most metal-poor bin ($-2.0 < \mathrm{[Fe/H]} < -1.5$) is omitted, as detectability is zero across its full extent. }  \label{fig:kiel_sensitivity_all}
\end{figure*}

Fig.~\ref{fig:kiel_sensitivity_all} illustrates the Ce sensitivity across almost the full atmospheric parameter space covered by the grid, for $-0.45<$~[Ce/Fe]~$<+0.95$. As discussed in Sect.~\ref{sect:snr}, the
sensitivity increases with both Ce abundance and metallicity, and satisfies
the detection threshold mainly for cool, near-solar-metallicity giants. 

Even at the highest Ce enhancements, for $\rm{SNR}<200$ detectability is never reached for 
\logg~$\gtrsim2.0$, where the Ce line remains too weak to be measured. This implies that Ce measurements in this region of parameter space should therefore be treated
with caution, since a non-detection reflects insufficient sensitivity rather than a genuine absence of Ce.

\section{Normalization effect}\label{append:norm}

Typical uncertainties in local, line-by-line normalization, particularly for more line-crowded and cooler stars, range from $\sim 0.2\%$ for high-SNR optical spectra to $\sim 2\%$, corresponding to abundance errors of 0.1-0.2 dex \citep[e.g.,][]{2020Cretignier}. We therefore consider $\epsilon = 0.2\%, 1\%$, and $2\%$.
Because the model flux is locally linear in [Ce/Fe], as assumed in the CRB, a constant offset $\epsilon$ is, to first order, equivalent to an abundance shift producing the same flux variation. We derive this shift below.

By adopting the same local linearity assumption that underlies the Cramér-Rao analysis, the continuum-normalized flux can be approximated around a reference abundance as:
\begin{equation}
f(\lambda,[\mathrm{Ce/Fe}]) \approx
f_0(\lambda)
+
\frac{\partial f}{\partial[\mathrm{Ce/Fe}]}
\,\Delta[\mathrm{Ce/Fe}],
\end{equation}

where $f_0(\lambda)
$ is the synthetic spectrum at the reference abundance 
$[\mathrm{Ce/Fe}]_{0}$. 

Now consider a small continuum-normalization error represented by a fractional offset $\epsilon$. Since the continuum-normalized flux is close to unity ($f_0(\lambda)\simeq1$) near the continuum, the observed normalized flux can be approximated as:
\begin{equation}
f_{\rm obs}(\lambda)
=
f_0(\lambda)(1+\epsilon)
\simeq
f_0(\lambda)+\epsilon,
\end{equation}
Since the fitting procedure assumes the continuum is correctly normalized,
it attributes the flux offset to a change in abundance, yielding:
\begin{equation}
\epsilon
\approx
\frac{\partial f}{\partial[\mathrm{Ce/Fe}]}
\,\Delta[\mathrm{Ce/Fe}]_{\rm ind},
\end{equation}
where $\Delta[\mathrm{Ce/Fe}]_{\rm ind}$ denotes the abundance shift induced by the continuum-placement error, giving:
\begin{equation}
\Delta[\mathrm{Ce/Fe}]_{\rm ind}
\approx
\frac{\epsilon}
{\partial f/\partial[\mathrm{Ce/Fe}]}.
\end{equation}
Since we adopt the peak gradient, the resulting bias represents a conservative lower limit. 

\begin{figure*}[t]
    \centering   \includegraphics[width=0.8\textwidth]{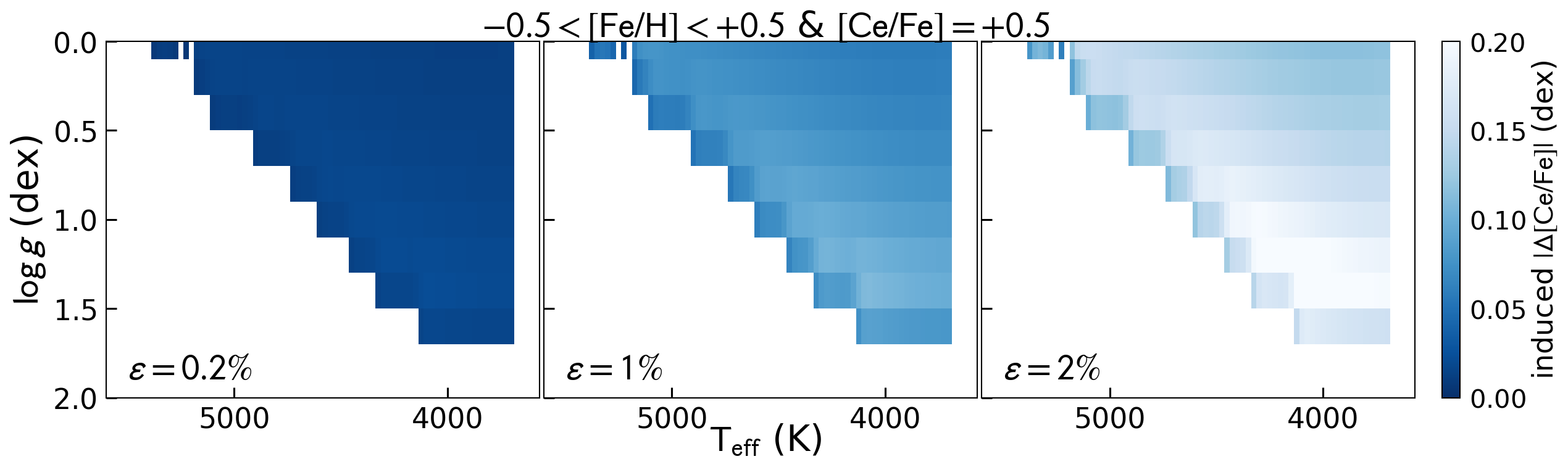}
    \caption{Sensitivity to continuum-normalization errors for three continuum-error levels ($\epsilon$), quantified by the induced shift in [Ce/Fe]. The color scale indicates the magnitude of this shift, with darker blue denoting a smaller induced bias (and therefore more reliable abundance recovery) and lighter shades denoting a larger bias, consistent with the colour convention used throughout the paper. White regions mark where the line is undetectable according to the combined EW and CRB criterion.} \label{fig:continuum}
\end{figure*}

Figure~\ref{fig:continuum} shows the induced bias across the Kiel diagram for $-0.5<\mathrm{[Fe/H]}<+0.5$ and [Ce/Fe]~$=+0.5$, restricted to the region where Ce is detectable under our combined EW and CRB criterion. Within this region, the induced continuum bias remains below our 0.1 dex threshold for $\epsilon=0.2\%$ (maximum bias $\approx0.03$ dex). At $\epsilon=1\%$, it exceeds 0.1 dex in $20.2\%$ of the detectable region, although it already reaches substantial values of $0.06$–$0.11$ dex elsewhere. For the largest error considered ($\epsilon=2\%$), the bias exceeds 0.1 dex throughout the entirety of the detectable region (minimum $\approx0.112$ dex). Continuum-placement errors therefore become the dominant source of uncertainty for $\epsilon\gtrsim1\%$. 

Finally, our continuum-suppression analysis implicitly assesses systematics arising from the metallicity-dependent $\alpha$-enhancement of real stars and the CN molecular feature next to the Ce triplet, which is deliberately excluded from our gradient window to isolate the Ce signal. This is particularly relevant for s-process-enhanced stars (e.g., barium and CEMP-s stars), in which C and Ce are expected to increase concomitantly due to AGB nucleosynthesis \citep{2011Bisterzo}, strengthening the CN blend precisely where the Ce feature is most prominent. Our synthesis grid adopts solar-scaled C and N abundances, so this correlated enhancement is not modelled explicitly. However, the CN feature's primary effect, suppression of the local pseudo-continuum, is encompassed by our continuum-bias analysis, which explores a range of $\epsilon$ values. A similar effect may arise from TiO molecular transitions. However, a comparison of synthetic spectra computed with and without TiO line opacity at \Teff$=4500$ and $3900$~K shows that the mean change in normalised flux across the Ce window between the two temperatures is $\sim$0.01 ($\sim$1\% of the continuum) in both cases, i.e. the inclusion of TiO lines does not increase it. Any residual TiO contribution therefore acts as a smooth suppression of the local pseudo-continuum, which is also accounted for by the range of $\epsilon$ values considered here. 

Finally, as molecular absorptions strengthen at lower temperatures, Ce detectability may be further reduced in the coolest stars (\Teff$\lesssim3900$~K), making our limits optimistic.

\section{Comparison with the Gaia \texttt{CeUpLim} Classification}\label{append:E}
\begin{table*}[ht!]
\centering
\caption{Fraction of Gaia DR3 stars with SNR$\geq50$ and a measured Ce
abundance (\texttt{CeUpLim}$\leq2$) falling inside/outside the theoretical
Ce detectability region, by \texttt{CeUpLim} tier and metallicity bin.}
\label{tab:ceuplim_stats}
\begin{tabular}{llrrrrr}
\hline \hline
[Fe/H] & \texttt{CeUpLim} & N & Inside (EW+CRB) & Outside (EW+CRB) & Inside (EW+CRB+$\epsilon$) & Outside (EW+CRB+$\epsilon$) \\
\hline
$-1.5$ to $-0.5$ & 0 & 3    & 33.3 & 66.7 & 0.0  & 100.0 \\
$-1.5$ to $-0.5$ & 1 & 6    & 16.7 & 83.3 & 0.0  & 100.0 \\
$-1.5$ to $-0.5$ & 2 & 6    & 16.7 & 83.3 & 0.0  & 100.0 \\
$-0.5$ to $0.0$  & 0 & 573  & 69.5 & 30.5 & 34.7 & 65.3 \\
$-0.5$ to $0.0$  & 1 & 1953 & 67.4 & 32.6 & 34.6 & 65.4 \\
$-0.5$ to $0.0$  & 2 & 5247 & 62.7 & 37.3 & 36.5 & 63.5 \\
$0.0$ to $0.5$   & 0 & 80   & 93.8 & 6.2  & 72.5 & 27.5 \\
$0.0$ to $0.5$   & 1 & 325  & 99.1 & 0.9  & 64.9 & 35.1 \\
$0.0$ to $0.5$   & 2 & 1546 & 99.0 & 1.0  & 48.7 & 51.3 \\
\hline \hline
\end{tabular}
\tablefoot{Columns 4-5 ("EW+CRB") give the fraction of stars whose measured \Teff\ and \logg, evaluated at the detectability boundary of the nearest of our three reference [Ce/Fe] values ($-0.5$, $0.0$, $+0.5$; as in Sect.~\ref{sect:comparison}), fall inside/outside our combined equivalent-width and Cramér-Rao criterion at SNR$=50$ (Sect.~\ref{sect:detectability}). Columns 6-7 additionally require the induced continuum bias at $\epsilon=1\%$ to remain below our $0.1$~dex threshold (Sect.~\ref{sect:continuum}).}
\end{table*}

Table~\ref{tab:ceuplim_stats} extends the comparison of Sect.~\ref{sect:comparison} by reporting, for each metallicity bin and \texttt{CeUpLim} tier, the fraction of Gaia DR3 stars (SNR~$\geq50$) whose own measured atmospheric parameters and [Ce/Fe] fall inside our detectability criterion. Of 9930 stars considered in Sect.~\ref{sect:comparison}, 9739 fall within the three metallicity bins tabulated here (the remainder lying outside $-1.5\leq\mathrm{[Fe/H]}\leq0.5$ or lacking a metallicity estimate).

The metal-poor bin ($-1.5\leq\mathrm{[Fe/H]}<-0.5$) contains only 3--6 stars per tier and is statistics-limited; no robust conclusion can be drawn from it beyond the qualitative consistency with the near-null detectability found in Sect.~\ref{sect:snr}. For $-0.5\leq$~[Fe/H]~$<0.0$, $63$--$69\%$ of stars fall inside our boundary under the EW+CRB criterion, but this drops to $\sim35\%$ once the $\epsilon=1\%$ continuum bias is applied, underscoring the importance of continuum placement in this regime. The metal-rich bin ($0.0\leq\mathrm{[Fe/H]}\leq0.5$) is almost entirely enclosed by the EW+CRB boundary ($94$--$99\%$), but the continuum bias again reduces this substantially, to $49$--$73\%$. Notably, this reduction is strongest for the lowest-confidence tier (\texttt{CeUpLim}$=2$), falling to $\sim49\%$, and weakest for the best-quality measurements (\texttt{CeUpLim}$=0$), $\sim73\%$. This suggests that, once continuum placement is accounted for, the \texttt{CeUpLim} flag and our theoretical criterion capture related but not identical aspects of measurement reliability.

\end{appendix}

\end{document}